\documentclass[a4paper,11pt]{article}

\usepackage{jcappub}
\usepackage{newtxtext,newtxmath}

\usepackage[T1]{fontenc}
\usepackage{ae,aecompl}
\usepackage{booktabs}

\usepackage{graphicx}	
\usepackage{amsmath}	
\usepackage{amssymb}	
\usepackage{xcolor}
\usepackage{bm}
\usepackage{color}

\title{Impact of Polarized Galactic Foreground Emission on CMB Lensing Reconstruction and Delensing of B-Modes}

\author[a]{Dominic Beck}
\author[a]{Josquin Errard}
\author[a,b]{Radek Stompor}

\affiliation[a]{AstroParticule et Cosmologie, Univ Paris Diderot, CNRS/IN2P3, CEA/Irfu, Obs de Paris, Sorbonne Paris Cit\'e, France}
\affiliation[b]{CNRS-UCB International Research Laboratory, Centre Pierre Bin\'etruy, UMI2007}

\emailAdd{dbeck@apc.in2p3.fr}

\abstract{
Next generation CMB experiments such as CMB-S4 aim at measuring the CMB lensing potential at sub-percent precision where most of the constraining power will come from CMB polarization. We investigate the prospects of achieving this goal in the presence of large-scale, diffuse galactic foreground emission by using non-Gaussian 
sky simulations and exploit multi-frequency information to clean those. We show that, while prior to foreground cleaning, cosmological parameter estimates from the contaminated lensing potential estimation can be significantly biased, these can be successfully mitigated by applying a parametric foreground cleaning approach. We further observe no significant additional bias in the delensed B-mode power spectrum after applying foreground cleaning and are therefore able to obtain an unbiased measurement of the tensor-to-scalar ratio, $r$, after delensing.
}

\begin{document}
\label{firstpage}
\maketitle
\flushbottom

\section{Introduction}
\label{sec:intro}
Following the conclusion of the Planck mission \citep{planck2018:overview}, which was able to measure the cosmic microwave background (CMB) intensity power spectrum close to its cosmic-variance limited sensitivity down to scales of a few arcminutes, we enter an era of precision measurements of CMB polarization. Future CMB experiments are planning to observe the polarized CMB with unprecedented sensitivity \citep{Abazajian2016}. Major science targets of these experiments include a measurement of large-scale B-mode polarization, in particular the tensor-to-scalar ratio, $r$, determining the amplitude of the primordial gravitational wave power spectrum \citep{Kamionkowski1996,sz97}, as well as the dark-sector and neutrino physics imprinted in the small-scale E- and B-mode power spectra and the CMB lensing potential power spectrum. \\

The lensing potential is the line-of-sight-integrated gravitational potential and its gradient describes the deflection of the primary CMB photons' paths by the inhomogeneous matter distribution in the Universe. It can be estimated from the CMB itself from quadratic combinations of the CMB intensity, $I$, and polarization, $E$ and $B$ \citep{Okamoto2003}. In noise regimes where the small-scale B-mode polarization becomes signal dominated, the lensing potential estimation using polarization estimators, in particular from the $E-B$ correlation, starts dominating in sensitivity over the estimation from CMB temperature, which delivered the most powerful lensing potential measurement to date \citep{planck-lensing2018,Wu2019}. For temperature lensing reconstruction, several astrophysical contaminants have been identified and treated, like the emission of the cosmic infrared background and radio point sources \citep{VanEngelen2014,Osborne2014}, the kinematic Sunyaev-Zeldovich (kSZ) effect \citep{FerraroHill2018} and the thermal Sunyaev-Zeldovich (tSZ) effect \citep{Madhavacheril2018}. These effects were found to be less severe or non existent in the CMB polarization channels \citep{Stein1966,Hildebrand1988,Feng2019,Smith2009,Sazonov1999,Hall2014a}. Besides extragalactic foregrounds, diffuse galactic foreground emission due to thermal dust and synchrotron radiation are significant foregrounds to CMB observations in both temperature and polarization. Their effect on CMB lensing reconstruction has been investigated in Ref.~\citep{Fantaye2012} based on the knowledge of galactic foregrounds we had prior to the Planck mission. Since then, thanks to high signal-to-noise measurements of the galactic emission intensity as well as the full-sky polarisation maps in a broad range of frequencies, we gained deeper insight into the complexity of galactic foreground emission, especially in its polarization properties. This includes the variation of polarization fraction \citep{PlanckDustSED} and of spectral indices of the SEDs \citep{Planck2018dust,Krachmalnicoff2018} across the sky. Several mitigation strategies of diffuse foreground contamination specifically in lensing potential estimates have been proposed in the literature. This includes the cleaning of only one leg of the quadratic estimator \citep{Madhavacheril2018}, the so-called shear estimator, which is more robust to foreground biases \citep{Schaan2018a} and the bias-hardened quadratic estimator against a varying dust amplitude with given power spectrum \citep{planck-lensing2018}. \\

CMB lensing requires understanding of the small-scale, non-Gaussian anisotropies that are present in the measured CMB maps. With increasing sensitivity in CMB lensing potential measurements, the foreground models have to capture the non-Gaussian statistic to an appropriate level. Only with the current onset of multi-frequency, high-sensitivity observations from the ground, we will be able to improve our understanding of galactic foregrounds to the CMB beyond multipoles $\ell \approx 300$, where the polarized dust power spectrum from Planck is signal dominated at high galactic latitudes. Future multi-frequency observations and component separation, however, are not fail-safe systems. We know that, given the possible complexity of polarized galactic foregrounds, neither avoiding regions with high galactic foreground amplitude in the survey~\citep{Planck2018dust,Krachmalnicoff2015,Krachmalnicoff2018} nor component separation techniques (e.g. in Refs.~\citep{Errard2016,Errard2018}) can assure an unbiased estimate of a primordial gravitational wave signal in the B-mode power spectrum. Since future CMB lensing potential power spectrum measurements will be driven by measuring the non-Gaussian correlations in the CMB polarization, those measurements, and subsequent parameter estimations, might be biased, even after attempting multi-frequency galactic foreground cleaning. Furthermore, any higher-order correlations in the galactic foregrounds between small- and large-scales might introduce unexpected biases in the delensed B-mode power spectrum and hence bias a tensor-to-scalar ratio measurement. Those higher-order correlations of foregrounds have already been detected in data \citep{Jung2018,Coulton2019}.\\

In this paper we address the impact of galactic diffuse foregrounds in the measured CMB on the reconstructed CMB lensing potential with the quadratic estimator and the subsequent CMB lensing power spectrum measurement and delensed CMB B-mode power spectrum. We report biases and sensitivity degradation in parameter estimates of the tensor-to-scalar ratio, $r$, and total mass of neutrinos, $M_\nu$, caused by statistical foreground models relying on simulations of the galactic magnetic field (GMF). We start by reviewing the Galactic foreground simulations we use in this work, continue by assessing the component separation performance for the future ground-based CMB experiment CMB-S4 and characterize foreground biases after component separation. Next, we will characterize biases in the CMB lensing potential and delensed B-mode power spectra, before propagating it to the two major cosmological parameters of interest, $r$ and $M_\nu$.

\section{Methods}
\label{sec:method}

In this section we summarize the simulations and methods that we use to clean the foregrounds as well as estimate the lensing potential power spectrum and delensed B-mode power spectrum.

\subsection{Sky Components}

We model (single-frequency or foreground-cleaned) $I$, $Q$ and $U$ maps as a sum of lensed CMB, $\mathbf{s}$, instrumental noise, $\mathbf{n}$, and an additional galactic foreground component, $\mathbf{f}$
\begin{equation}
    \mathbf{d}\equiv(I,Q,U)^T=\mathbf{s}+\mathbf{n}+\mathbf{f}.
    \label{componentdefinition}
\end{equation}
We denote the angular power spectrum of these fields with $C_\ell^{XY}$, where $X,Y \in \{T,E,B\}$, defined by
\begin{equation}
    C_\ell^{XY}\equiv\frac{1}{2\ell+1}\sum_m a^{X \dagger}_{\ell m} a^Y_{\ell m} ,
    \label{eq:powspecdefinition}
\end{equation}
given the harmonic coefficients $a^X_{\ell m}$ of the $I$, $Q$ and $U$ maps. 

\subsubsection{CMB}

In our simulations, the CMB is generated via random Gaussian realizations of unlensed $I$, $Q$ and $U$ maps, as well as the lensing potential $\phi$ from power spectra computed with {\tt CAMB} given the \textsc{Planck} {\tt TT,TE,EE+lowP+lensing+ext} best-fit cosmological model \citep{Ade2016a}. We then produce lensed CMB maps with {\sc lenS2HAT} \citep{fabbian2013}. 

\subsubsection{Noise and observation strategy}

\begin{figure}
\centering
\includegraphics[width=.33\linewidth]{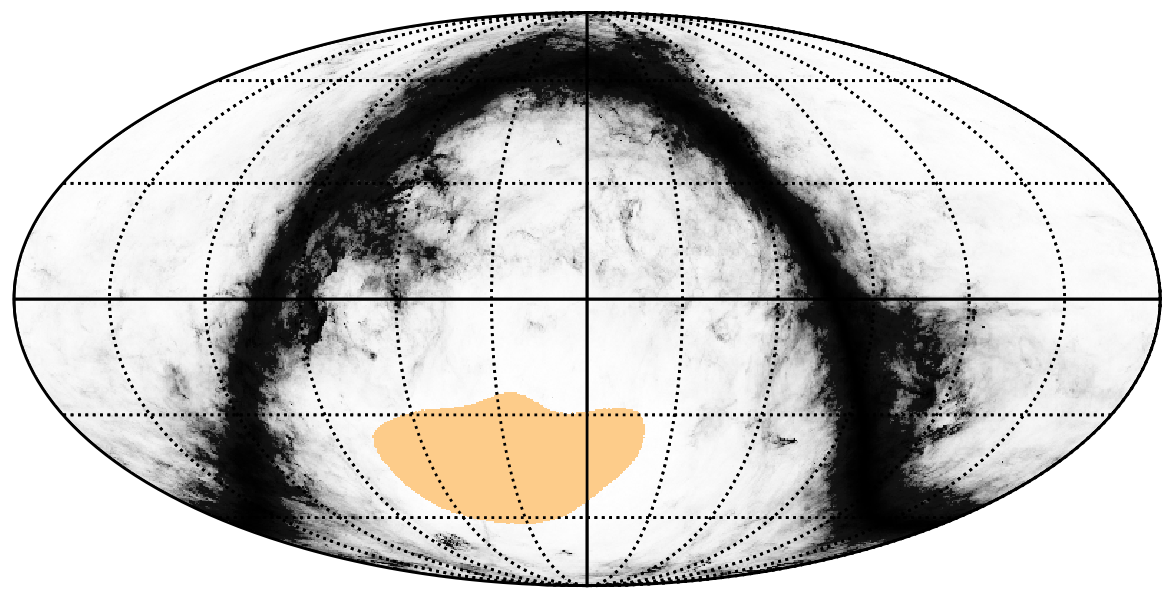}\hfill
\includegraphics[width=.33\linewidth]{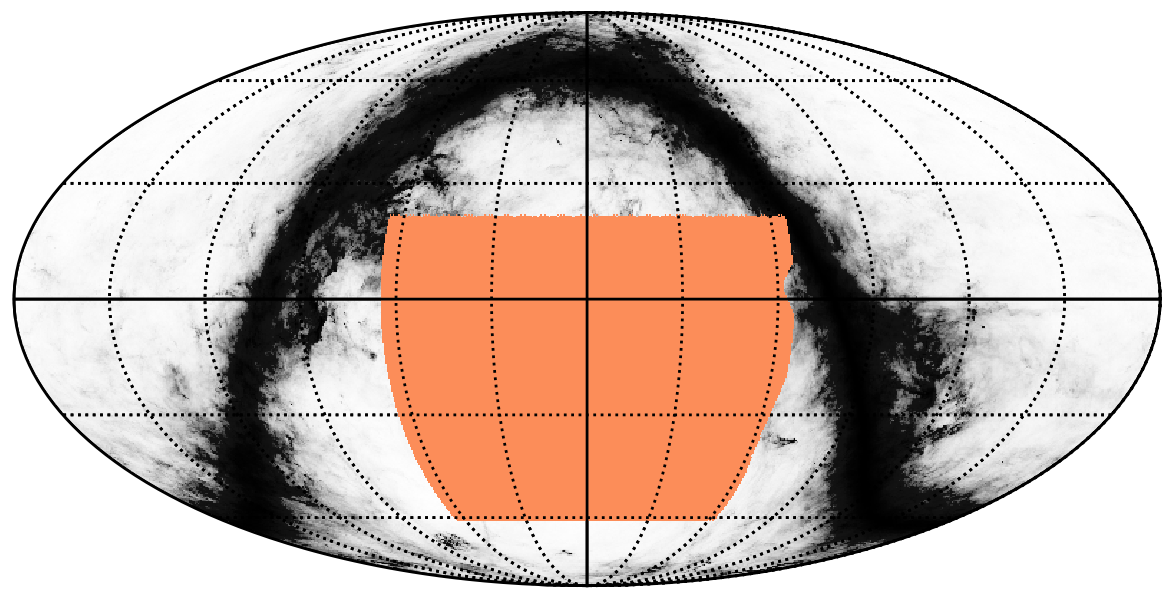}\hfill
\includegraphics[width=.33\linewidth]{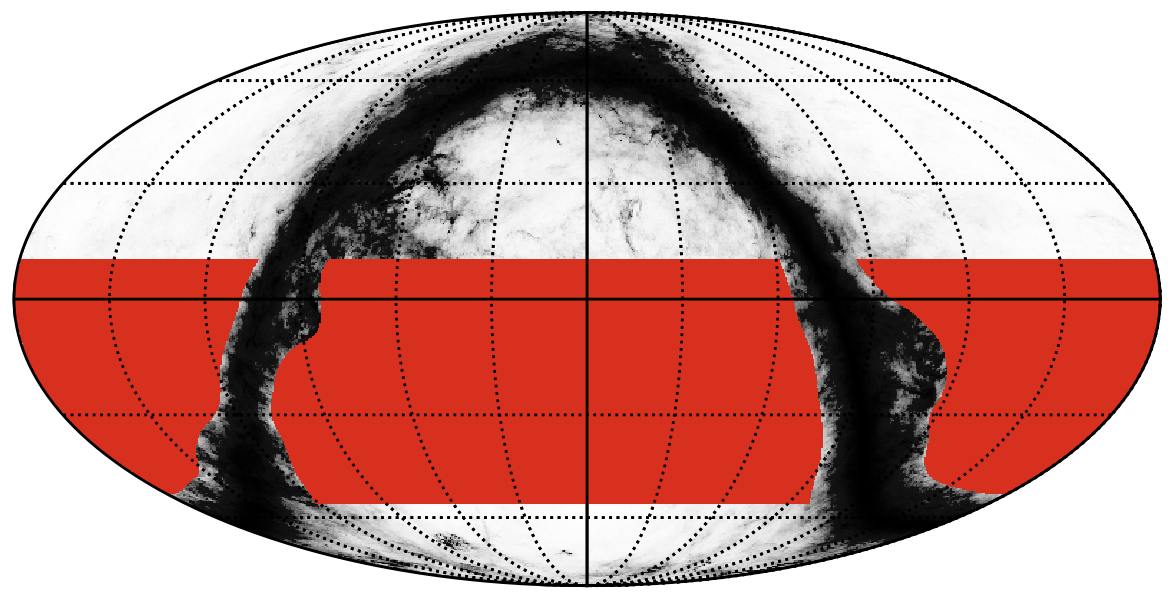}
\caption{The considered survey footprints, spanning (from left to right) $5$\%, $22$\% and $40$\% of the sky, shown in shades of red in Mollweide projection and celestial coordinates. The galactic dust intensity template of {\sc PySM} is shown in black.}
\label{fig:scansurveys}
\end{figure}

We assume a realistic experimental high-resolution configuration for future ground-based experiments, populating all available atmospheric windows in the microwave spectrum between 20 and 270 GHz with detectors observing the sky in seven frequency channels. As an observational strategy we investigate potential sky fractions, which are accessible from the Atacama desert in Chile. This includes a small, wide and ultra-wide survey strategy, covering about 5\%, 22\% and 40\% of the sky, respectively. Their footprints are shown in Fig.~\ref{fig:scansurveys}. To simulate noise with realistic power spectra we make use of the publicly availabe CMB-S4 noise calculator\footnote{\url{https://cmb-s4.org/wiki/images/Lat-noise-181002.pdf}}. Corresponding white noise levels and Gaussian beam widths are given in Table~\ref{tab:sensitivities}. \\

\begin{figure}
\centering
\includegraphics[width=.66\linewidth]{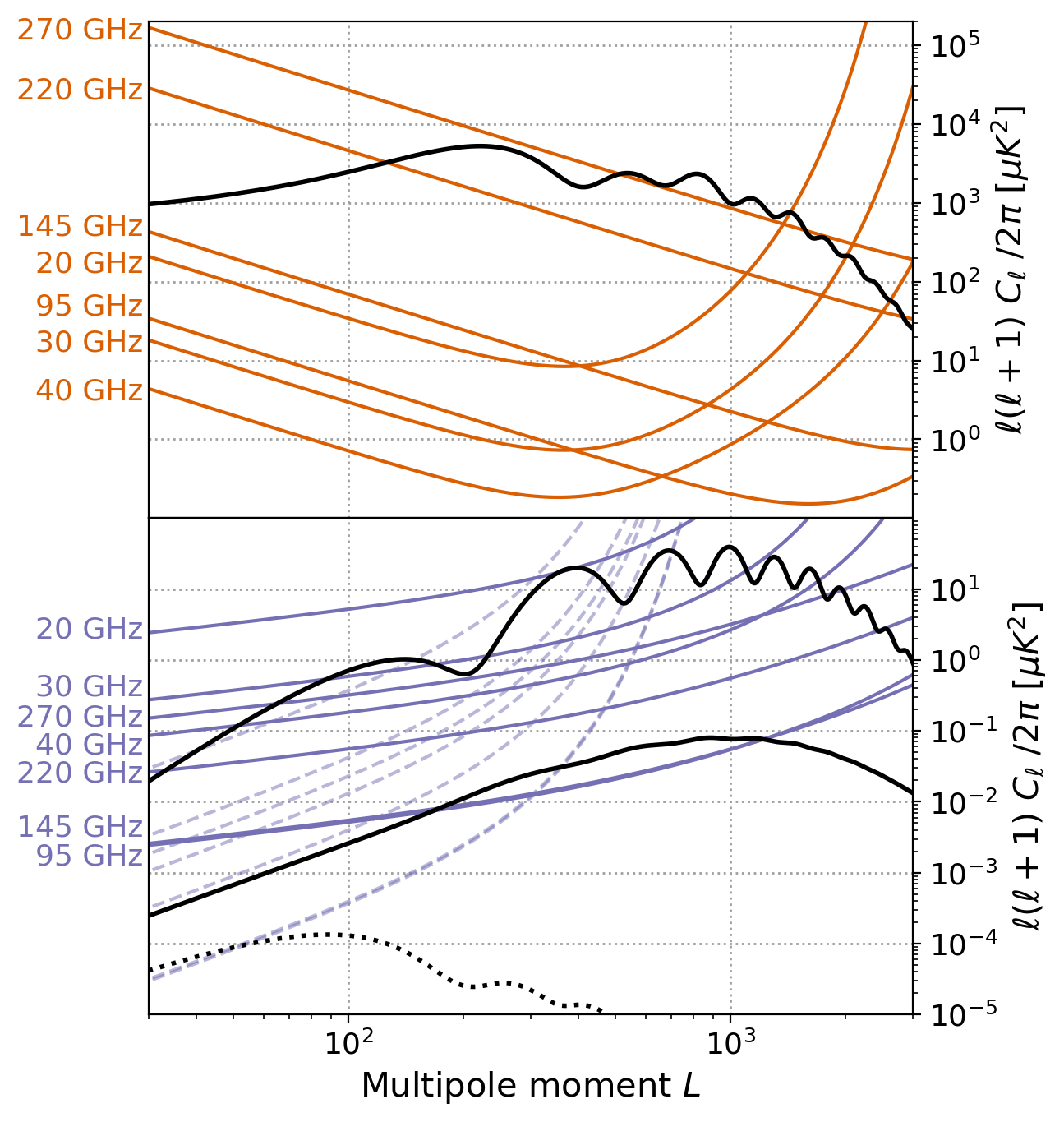}
\caption{The temperature (top) and polarisation (bottom) beam-deconvolved noise power-spectra for each of the seven frequency channels. The dashed lines in the bottom plot are the corresponding noise curves without the atmospheric component at lower resolution, to mimic a small-aperture telescope configuration targeting the large-scale B-mode spectrum. The thick black lines are the fiducial TT, EE and BB power-spectra, in that order from top to bottom. The target, the primordial B-mode power-spectrum corresponding to $r=10^{-3}$, is shown as the dotted black line.}
\label{fig:latnoisecurves}
\end{figure}

Additional to the large-aperture telescope (LAT) configuration targeting the small-scale CMB for lensing science, we assume a small-aperture telescope (SAT) setup, which is capable of observing atmosphere-free polarized CMB down to multipoles of $\ell=30$ in the same frequency channels with similar white noise levels to allow for the cleaning of galactic foregrounds. This could be achieved from the ground with a continuously-rotating half-wave plate \citep{Takakura2017}. We don't consider temperature information from the SAT. The assumed combined multipole range for both SAT and LAT stretch from $\ell=3000$ down to  $\ell=30$, if not stated otherwise. In Fig~\ref{fig:latnoisecurves} we show the corresponding beam-deconcolved noise curves for our SAT and LAT configurations. In the following we denote the noise power spectra with $N_\ell$ and the harmonic multipoles of its Gaussian realizations with $n_{\ell m}$. 

{
\setlength{\tabcolsep}{1mm}
\begin{table}
\centering
\begin{tabular}{r|cccccccc} \toprule
\textbf{frequency [GHz]} & \textbf{20} & \textbf{30} & \textbf{40} & \textbf{95} & \textbf{145} &  \textbf{220} & \textbf{270} \\ 
\hline
$\sigma_P$ [$\mu K$-arcmin] & 48.1 & 16.2 & 9.07 & 1.53 &  1.57 & 5.01 & 12.0
  \\
$\theta$ [arcmin] & 10 & 7.4 & 5.1 & 2.2 & 1.4 & 1.0 & 0.90 \\
\bottomrule
\end{tabular}
\caption{Frequency channels and corresponding instrumental sensitivity in terms of the white noise level in polarisation $\sigma_P$ and beam size $\theta$ for the considered CMB surveys.}
\label{tab:sensitivities}
\end{table}
}

\subsubsection{Templates of polarized galactic foregrounds}\label{sec:galmodels}

Polarized galactic foregrounds pose a major difficulty for future high-precision CMB polarization measurements. Fortunately, a subset of galactic components which are present in the temperature CMB data is believed to be not or only weakly polarized, such as the cosmic infrared background (CIB), bremsstrahlung (free--free) or anomalous microwave emission (AME) \citep{Stein1966,Hildebrand1988,Feng2019,Macellari2011,Rubino-Martin2012}. The perils of large-scale foregrounds from galactic polarized dust and synchrotron emission in the polarization power spectra are well identified as a major obstacle in the search for primordial gravitational waves. Thanks to measurements from WMAP, Planck, C-BASS and S-PASS \citep{Bennett2012,Planck2018dust,Krachmalnicoff2015,Krachmalnicoff2018}, data of the large-scale contamination in the polarization power spectra is available and already used to characterize and subtract \citep{BK2015a,BK2018a} these foregrounds. Besides having an anisotropic polarization fraction up to 20\% over the sky \citep{PlanckDustSED}, galactic dust and synchrotron emission has a direction dependent spectral energy distribution (SED) \citep{Planck2018dust,Krachmalnicoff2018}, adding additional complexity to its spatial and frequency dependency.\\

Information about small-scale polarized galactic foreground, which is relevant for lensing science, however, is sparse and remains largely unknown. Hence we chose to rely for this paper on data-driven statistical simulations to obtain estimates of foreground biases in the lensing power spectrum and delensed B-mode power spectrum.

\paragraph{PySM foreground simulations}

A first set of simulations is based on the polarized dust and synchrotron templates of \textsc{PySM}\footnote{\url{https://github.com/bthorne93/PySM_public}} \citep{Thorne2017}. For synchrotron these are a combination of the $408\ \textrm{MHz}$ Haslam maps \citep{Haslam1982,Remazeilles2014} and WMAP 9-year $23\ \textrm{GHz}$ maps \citep{Bennett2012} smoothed to five degrees. For thermal dust the template is based on the \texttt{COMMANDER} astrophysical component maps of polarized galactic dust emission of the Planck 2015 data release (PR2) \citep{Planck2015fgmaps}, smoothed to a resolution of two degrees. \\

Small scales are added, similar to the processing in \S3 of Ref.~\citep{Thorne2017}, by fitting the power spectrum at large-scales with a power-law in $\ell$,
\begin{equation}
    C^{EE/BB}_\ell=A\left(\frac{\ell}{80}\right)^\alpha,
    \label{eq:fgpowerlaw}
\end{equation}
and generating a Gaussian random realization of $Q$ and $U$ Stokes parameters at small-scales, such that the sum of large- and small-scale maps have a power spectrum given by the earlier fitted power-law. The pure, cut-sky power spectra are computed with \texttt{X2Pure} \citep{Grain2009} on the patch given by the observing strategy we consider, using an analytical apodization with a tapering of $8~\textrm{deg}$.

\paragraph{Galactic magnetic field foreground simulations}\label{sec:galsim}

To simulate the diffuse polarized emission from galactic dust, the spin orientation of the dust grains has to be taken into account. To do so we follow the strategy introduced in Refs.~\cite{PlanckCollaboration2016,Vansyngel2016}. Their orientation roughly aligns perpendicularly with the galactic magnetic field (GMF) field lines~\citep{Davis1951}. We rely on a simulation of the three-dimensional GMF, whose structure is then projected on the two-dimensional sphere to be related to the measured $I$, $Q$ and $U$ Stokes parameters. We then express the GMF as the sum of its mean, $\mathbf{B}_0$, and turbulent component, $\mathbf{B}_t$,
\begin{equation}
    \mathbf{B}=\mathbf{B}_0+\mathbf{B}_t.
\end{equation}
Since we are interested in the contamination to the CMB observed in a small fraction of the sky at high galactic latitudes, we ignore galaxy-wide variations and hence assume that $\mathbf{B}_0$ has a fixed orientation, $(l_0,b_0)$, representing the mean orientation of the GMF in the solar neighbourhood. We follow directly Refs.~\citep{PlanckCollaboration2016,Vansyngel2016} and compute the turbulent component of the GMF in the direction $\mathbf{x}$ as
\begin{equation}
    \mathbf{B}_t (\mathbf{x})= \left| \mathbf{B}_0(\mathbf{x}) \right| f_M \hat{\mathbf{B}}_t(\mathbf{x}),
\end{equation}
where each component of $\mathbf{B}_t$ is a Gaussian realization with an angular power spectrum $C_\ell\propto\ell^{\alpha_M}$.

The Stokes parameters of an optically thin emission at frequency $\nu$ are then given by \citep{PlanckCollaboration2016,Vansyngel2016} 
\begin{align}
    I_\nu (\mathbf{x})&= \int B_\nu \left( T_d (\mathbf{x}) \right) \left[ 1-p_0 \left( \cos^2 \gamma (\mathbf{x}) - \frac{2}{3} \right) \right] d\tau_\nu (\mathbf{x}) \\
    Q_\nu (\mathbf{x})&= \int p_0 B_\nu \left( T_d (\mathbf{x}) \right) \cos \left( 2\phi (\mathbf{x}) \right) \cos^2 \gamma (\mathbf{x}) d\tau_\nu (\mathbf{x}) \\
    U_\nu (\mathbf{x})&= \int p_0 B_\nu \left( T_d (\mathbf{x}) \right) \sin \left( 2\phi (\mathbf{x}) \right) \cos^2 \gamma (\mathbf{x}) d\tau_\nu (\mathbf{x}),
    \label{eq:foregroundmodel}
\end{align}
where $B_\nu \left( T_d \right)$ is the spectrum of a black body given by
\begin{equation}
    B_\nu(T_d)=\frac{2 h \nu^3 / c^2}{\textrm{exp}\left(\frac{h \nu}{k T_d}\right)-1},
\end{equation}
with dust temperature $T_d$, and the polarization fraction parameter, $p_0$. We denote with $\gamma$ the angle between the local magnetic field and the plane perpendicular to the line-of-sight, $\phi$ is the local polarization angle in the {\sc HEALPix}\footnote{\url{http://healpix.sourceforge.net/}} \citep{Gorski2005} convention and $\tau_\nu$ is the optical depth. We discuss the scaling laws of the foregrounds in the next section. As in Ref.~\citep{Vansyngel2016} we approximate the integrals along the line-of-sight with finite sums over source functions in $N$ layers of dust. The model of polarized galactic dust emission hence comprises of six parameters, $l_0$, $b_0$, $p_0$, $N$, $\alpha_M$ and $f_M$, which can be fitted against observed spectra at large angular scales by Planck \citep{Vansyngel2016}.\\

In practice we use the thermal dust intensity template of the {\sc PySM} package, which uses a realization of the extrapolated power spectrum below scales of $\approx 7$ arcmin, as a tracer of the source functions for the $Q$ and $U$ Stokes parameters. We also test directly the simulations produced as described in \citep{Vansyngel2016}, which furthermore features simulated $E-B$ power asymmetry. The $TE$ correlation is neglected. \\

Due to effects related to the projection from the three-dimensional GMF simulation to the two-dimensional polarization angle field, the resulting statistics of small-scale polarization is not anymore Gaussian and hence expected to affect measurements of higher-order correlations of the CMB, such as the CMB lensing potential or delensed B-mode spectra. 

\subsubsection{Frequency scaling}\label{sec:freqscalingsims}

The procedure described in the last section results in two sets of templates, including polarized thermal dust emission at $\nu_\textrm{dust}=353\ \textrm{GHz}$, $Q^\textrm{dust} (\mathbf{x})$ \& $U^\textrm{dust} (\mathbf{x})$, and polarized synchrotron emission at $\nu_\textrm{sync}=23\ \textrm{GHz}$, $Q^\textrm{sync} (\mathbf{x})$ \& $U^\textrm{sync} (\mathbf{x})$. These are used to simulate multi-frequency maps using \textsc{PySM} and following the explicit scaling factors 
\begin{align}
    A^{\nu\ \textrm{dust}}&=\left(\frac{\nu}{\nu_\textrm{dust}} \right)^{\beta_d-2} \frac{B_\nu(T_d)}{B_{\nu_\textrm{dust}}(T_d)}\\
    A^{\nu\ \textrm{sync}}&=\left(\frac{\nu}{\nu_\textrm{sync}} \right)^{\beta_s},
\end{align}
where $\beta_d$ and $\beta_s$ are the dust and synchrotron spectral index, $T_d$ the dust temperature.\\

The (noise-free) data model of an observed map at frequency $\nu$, assuming delta-shaped band-passes, is then given by
\begin{equation}
    \tilde{I} (\mathbf{x},\nu) = I^\textrm{CMB} (\mathbf{x}) + A^{\nu\ \textrm{dust}} (\mathbf{x}) \times I^\textrm{dust} (\mathbf{x}) + A^{\nu\ \textrm{sync}}(\mathbf{x}) \times I^\textrm{sync} (\mathbf{x}),\notag
\end{equation}
and equivalent expressions for $Q$ and $U$, assuming independent spectral indices between intensity and polarization. Here we explicitly allow for the frequency scaling factors to vary over the sky, corresponding to spatially varying spectral indices $\beta_d$ and $\beta_s$ \footnote{We assume a spatially constant dust temperature throughout this chapter.}, a fact established but weakly constrained by observations \citep{Planck2018dust,Krachmalnicoff2018}. In the simulations we either use constant spectral indices or spectral indices maps from Planck's \textsc{COMMANDER} pipeline \citep{Planck2015fgmaps} for dust and a combination of Haslam and WMAP maps for synchrotron, provided within \textsc{PySM} \citep{Thorne2017}.

\subsection{Foreground cleaning}\label{sec:fgresidualcompuatation}

The basis of our foreground cleaning algorithm is a maximum-likelihood solver with parametric foreground modeling, following Refs.~\citep{Stompor2009,Stompor2016}. The first step is the optimization of the spectral log-likelihood
\begin{align}
\mathcal{S}_\textrm{spec.}=-\int d^2\mathbf{x} \left(\mathbf{A}^T \mathbf{N}^{-1} \mathbf{d} \right)^T \left(\mathbf{A}^T \mathbf{N}^{-1} \mathbf{A} \right)^{-1} \mathbf{A}^T \mathbf{N}^{-1} \mathbf{d},
\end{align}
where $\mathbf{N}$ is the covariance of the noise, $\mathbf{n}$. This results in a set of estimates of spectral parameters, $\beta$, which can be used to reconstruct the mixing matrix $\hat{\mathbf{A}}=\hat{\mathbf{A}}(\mathbf{x},\beta)$. This estimate typically does not coincide with the true mixing matrix ${\mathbf{A}}(\mathbf{x})$, what leads leads to foreground residuals in the estimated, cleaned CMB map. An unbiased estimate of the components given a set of spectral parameters, $\beta$, can be written as 
\begin{align}
\hat{\mathbf{m}}(\mathbf{x})=\hat{\mathbf{W}}(\mathbf{x},\beta) \hat{\mathbf{d}}(\mathbf{x}),
\end{align}
with $\hat{\mathbf{W}} \equiv \left(\hat{\mathbf{A}}^T  \mathbf{N}^{-1} \hat{\mathbf{A}}  \right)^{-1} \hat{\mathbf{A}}^T\mathbf{N}^{-1}$. Given the true sky signal $\mathbf{s}$, we define residuals in the component maps as
\begin{align}
\mathbf{r}(\mathbf{x})=\hat{\mathbf{W}}(\mathbf{x},\beta) \hat{\mathbf{d}}(\mathbf{x})-\mathbf{s}(\mathbf{x})= \left(\hat{\mathbf{A}}^T  \mathbf{N}^{-1} \hat{\mathbf{A}}  \right)^{-1} \hat{\mathbf{A}}^T\mathbf{N}^{-1} \left( \mathbf{A} \mathbf{s}(\mathbf{x}) + \mathbf{n}(\mathbf{x}) \right) - \mathbf{s}(\mathbf{x})
\end{align}
In thermodynamical units we fix the scaling law of the CMB to be $\hat{\mathbf{A}}^\textrm{CMB}(\mathbf{x})=\mathbf{A}^\textrm{CMB}(\mathbf{x})=1$, for the assumed and the true sky. Using the same assumptions as in Ref.~\citep{Stompor2016}, this lets us write the foreground residuals in the reconstructed CMB map as
\begin{align}
\mathbf{r}^\textrm{CMB}(\mathbf{x})=\sum_k \mathbf{W}^{0k}(\mathbf{x},\beta) \hat{\mathbf{f}}^k(\mathbf{x}),
\label{eq:rescmb}
\end{align}
given foreground-only maps for each frequency channel $k$, $\hat{\mathbf{f}}_p^k$. Likewise, the noise after component separation can be constructed as
\begin{align}
\mathbf{n}^\textrm{CMB}(\mathbf{x})=\sum_k \mathbf{W}^{0k}(\mathbf{x},\beta) \hat{\mathbf{n}}^k(\mathbf{x}),
\end{align}
with map-space variance
\begin{align}
\left({\sigma^\textrm{CMB}}\right)^2(\mathbf{x})=\left[ \left(\hat{\mathbf{A}}^T(\mathbf{x}) \mathbf{N}^{-1}(\mathbf{x}) \hat{\mathbf{A}}(\mathbf{x}) \right)^{-1} \right]_{\textrm{CMB},\textrm{CMB}}
\end{align}
and power spectra
\begin{align}
N_\ell^\textrm{cleaned}=\left(\hat{\mathbf{A}}^T \mathbf{N}^{-1}_\ell \hat{\mathbf{A}}\right)^{-1},
\end{align}
assuming pixel-independent scaling laws. The implementation of the maximum-likelihood estimation and residual calculation makes extensive use of the {\sc fgbuster} library\footnote{\url{https://github.com/fgbuster/fgbuster}}. \\

Given the harmonic coefficients of the foreground signal, $f_{\ell m}$, the foreground-only power spectrum is obtained using Eq.~\ref{eq:powspecdefinition}. After component separation, the residual power from foregrounds is given by Eq.~\ref{eq:rescmb}. To account for both, the systematic bias due to mismatch between true and assumed model and statistical error due to the uncertainty in the model estimation caused by the ignorance of the spectral parameters, $\beta_i$, we expand the residual maps in powers of the spectral parameters, around the maximum likelihood estimate, $\hat{\beta}$,
\begin{align}
\mathbf{r}_p^{\rm CMB}=&\sum_k \mathbf{W}^{0k}_p (\hat{\beta}) \hat{\mathbf{f}}_p^k + \sum_{ki} \delta \beta_i \frac{\partial\mathbf{W}^{0k}_p}{\partial \beta_i} (\hat{\beta}) \hat{\mathbf{f}}_p^k + \sum_{kij} \delta \beta_i\delta \beta_j \frac{\partial^2\mathbf{W}^{0k}_p}{\partial \beta_i\partial \beta_j} (\bar{\beta}) \hat{\mathbf{f}}_p^k \\ \equiv &
\mathbf{y}+\sum_i \delta \beta_i \mathbf{Y}_i + \sum_{ij} \delta \beta_i \delta \beta_j \mathbf{X}_{ij}.
\end{align}
Here we adopt the notation of Ref.~\citep{Stompor2016}
\begin{align}
\mathbf{y}&\equiv \sum_k \mathbf{W}^{0k}_p (\hat{\beta}) \hat{\mathbf{f}}_p^k, \label{eq:defsmally} \\
\mathbf{Y}_i&\equiv \sum_{k} \frac{\partial\mathbf{W}^{0k}_p}{\partial \beta_i} (\hat{\beta}) \hat{\mathbf{f}}_p^k, \label{eq:defbigy}\\
\mathbf{X}_{ij}&\equiv \sum_{k} \frac{\partial^2\mathbf{W}^{0k}_p}{\partial \beta_i\partial \beta_j} (\hat{\beta}) \hat{\mathbf{f}}_p^k. \label{eq:defx}
\end{align}
With the spherical harmonic transforms of these quantities, we can write the (cross-)power spectra of foreground residuals in the cleaned CMB maps $X$ and $Y$ after averaging over noise realizations as \citep{Stompor2016}
\begin{align}
F_\ell^{XY~{\rm res.}}=&\frac{1}{2\ell+1} \sum_m y_{\ell m}^{X~\dagger} y_{\ell m}^Y +  y_{\ell m}^{X~\dagger} z_{\ell m}^Y +  z_{\ell m}^{X~\dagger} y_{\ell m}^Y +\notag \\&+ \sum_{ij} \boldsymbol{\Sigma}_{ij} \mathbf{Y}_{i\ell m}^{X~ \dagger} \mathbf{Y}_{j \ell m}^Y,
\label{eq:2ptfgres}
\end{align}
with 
\begin{equation}
    z_{\ell m}^X \equiv \sum_{ij} \Sigma_{ij} \mathbf{X}_{ij\ell m}^X
\end{equation}
and the covariance of the spectral parameters
\begin{equation}
    \Sigma_{ij}=\text{Cov}\left(\hat{\beta}_i,\hat{\beta}_j\right).
\end{equation}

\subsection{Lensing potential reconstruction}

For the estimation of the lensing potential, $\phi$, from the given CMB maps we employ \textsc{lensquest}\footnote{\url{https://github.com/doicbek/lensquest}}, a curved-sky implementation of the quadratic estimator of \cite{Okamoto2003} \footnote{The implementation follows closely \cite{Okamoto2003}, with the exceptions of neglecting the correlation between $I$ and $E$ and replacing the unlensed with the lensed CMB power spectra in the estimator weights following \citep{Hanson2011}}
\begin{align}
\hat{\phi}_{LM}^{XY}&=\sum_{\substack{\ell_1 m_1\\\ell_2 m_2}} \left[\frac{A^{XY}_L}{L(L+1)} (-1)^M
\begin{pmatrix}
\ell_1 & \ell_2 & L \\
m_1 & m_2 & -M
\end{pmatrix}
g^{XY}_{\ell_1\ell_2 L}\right] \hat{a}^X_{\ell_1 m_1} \hat{a}^Y_{\ell_2 m_2} \notag\\&\equiv \left[ \hat{a}^X_{\ell_1 m_1} \odot \hat{a}^Y_{\ell_2 m_2} \right]_{LM}. \label{eq:quadraticestimator}
\end{align}
For the definitions of $A^{XY}_L$ and $g^{XY}_{\ell_1\ell_2 L}$ the reader is referred to Ref.~\citep{Okamoto2003}. To forecast the capability of future CMB lensing measurements on large sky fractions of up to 40\% \citep{SO2018,Abazajian2016} and CMB scales as large as $\ell=30$ we work with curved-sky fields in {\sc HEALPix} pixelization. Prior to the reconstruction, we apply a tapered mask with a $8\ \textrm{deg}$ cosine-apodization to the input maps \citep{Grain2009}. \\

The $I$, $E$ and $B$ fields computed on the masked sky are used as inputs to the lensing estimator.
We follow Ref.~\citep{Benoit-Levy2013} to estimate the mean-field bias with the help of Monte-Carlo simulations and subtract it from the estimated field. To estimate the lensing potential power spectrum from the reconstructed lensing potential,  $\hat{\phi}_{LM}$, we compute the quadratic average over modes for a given $L$
\begin{equation}
\hat{C}_L^{\phi\phi}=W_4^{-1}\frac{1}{2 L +1} \sum_M \hat{\phi}^\dagger_{LM} \hat{\phi}^{\mathstrut}_{LM}-N_L,
\label{eq:lenspowerspec}
\end{equation}
where $W_4\equiv\int d \mathbf{x}\  m^4(\mathbf{x})$ and $m(\mathbf{x})$ is the apodized mask. $N_L$ is the sum of bias terms, in the following including the analytic computations of $N_L^{(0)}$ \citep{Okamoto2003} 
\begin{align}
    &N_L^{(0)~ABCD} \left[C_\ell\right] \equiv \frac{A_L^{AB}A_L^{CD}}{(L(L+1))^2 (2L+1)} \times \\&~~\times\sum_{\ell_1\ell_2} \left[ g^{AB\ \star}_{\ell_1\ell_2L}\left[ g^{CD}_{\ell_1\ell_2L} C_{\ell_1}^{AC}C_{\ell_2}^{BD}  + (-1)^{\ell_1+\ell_2+L}  g^{CD}_{\ell_2\ell_1L} C_{\ell_1}^{AD}C_{\ell_2}^{BC} \right] \right] \notag   
\end{align}
and $N_L^{(1)}$ \citep{Kesden2003}. Additional disconnected bias terms can be mitigated by choosing appropriate weights in the quadratic estimator \citep{Hanson2011}.\\

From Eqs.~\ref{componentdefinition}, \ref{eq:quadraticestimator} and \ref{eq:lenspowerspec} we obtain the bias terms including the contributions of the foreground components after averaging over the statistically isotropic components signal and noise, $\tilde{s}_{\ell m}+n_{\ell m}$, 
\begin{align}
    \left\langle \hat{C}_L^{\phi\phi} \right\rangle = C_L^{\phi\phi}+N_L^{(0)}\left[\hat{C}_\ell+F_\ell \right] + F_L^\textrm{syst.},
    \label{eq:fgsystbiasexpansion}
\end{align}
up to zeroth order in $C_L^{\phi\phi}$, where the contributions of the foregrounds to the power spectrum are included in the total measured power spectrum in the computation of $N_L^{(0)}$ and
\begin{equation}
F_L^\textrm{syst.} = W_4^{-1} \frac{1}{2L+1} \sum_M \left[ f \odot f \right]_{LM}^\dagger \left[ f \odot f \right]_{LM}^{\mathstrut} - N_L^{(0)} \left[ F_\ell \right].
    \label{eq:fgtrispectrum}
\end{equation}
We used here the notation introduced in Eq.~\ref{eq:quadraticestimator} to denote a quadratic combination of CMB fields to a lensing potential estimate with $\odot$. Eq.~\ref{eq:fgsystbiasexpansion} is true even when the foreground fields, $f_{\ell m}$, are non-isotropic. The $N_L^{(0)}$-terms including the two-point function of the foregrounds are naturally accounted for in a realization-dependent bias subtraction \citep{Namikawa2013}. The unknown trispectrum of the foregrounds in the first term of Eq.~\ref{eq:fgtrispectrum} is the potentially problematic one, as it will bias the lensing potential power spectrum estimate. In the presence of sky-masking and more complex noise than white noise, Eq.~\ref{eq:fgsystbiasexpansion} becomes inaccurate.\\

After component separation, similar to the case of the two point function, we can account for the uncertainty in the spectral parameters and obtain an expression for the resulting statistical bias in the lensing power spectrum. Using the definitions in Eqs.~\ref{eq:defsmally}, \ref{eq:defbigy} and \ref{eq:defx} in Eq.~\ref{eq:lenspowerspec}, we obtain
\begin{align}
    F_L^\textrm{stat.}&=W_4^{-1} \frac{1}{2L+1} \sum_M \left[ \mathbf{z} \odot \mathbf{y} \right]_{LM}^\dagger\left[ \mathbf{y} \odot \mathbf{y} \right]_{LM}^{\mathstrut} + \textrm{cycl.} + \\
    &+ W_4^{-1} \frac{1}{2L+1}  \sum_M\sum_{ij} \Sigma_{ij} \left[ \mathbf{Y}_i \odot \mathbf{Y}_j \right]_{LM}^\dagger \left[ \mathbf{y} \odot \mathbf{y} \right]_{LM}^{\mathstrut} + \textrm{perm.} \ \ .
    \label{eq:xfforlensing}
\end{align}
This amounts to $2+4 \times n+2 \times n^2$ quadratic estimator evaluations, where $n$ is the number of spectral parameters. Hence we can write the power spectrum of the reconstructed lensing potential to zeroth order as
\begin{equation}
    \left\langle \hat{C}_L^{\phi\phi} \right\rangle = C_L^{\phi\phi}+N_L^{(0)} \left[\hat{C}_\ell+F_\ell \right] +F_L^\textrm{syst.}+ F_L^\textrm{stat.}+... ,
    \notag
\end{equation}
where the last term is zero before component separation and $F^\textrm{syst.}$ needs to be computed using foregrounds residuals instead of full-power-input foregrounds, substituting $\mathbf{f}$ with $\mathbf{y}$ in Eq.~\ref{eq:fgtrispectrum}.

\subsection{Constructing the B-Mode Template and Delensing}

We follow Ref.~\citep{Smith2012} in creating a template of lensing-induced B-modes that can be used to subtract from the measured, total data to reduce the lensing variance in the final B-mode power spectrum. Given an estimated lensing potential with corresponding noise power spectrum, $N_L$, and the measured E-mode field as an approximation of the primordial E-mode field, we can write the B-mode template as
\begin{align}
    \hat{a}^{B~\textrm{temp.}~XY}_{\ell m} = &\sum_{LM\ell'm'} 
    \begin{pmatrix}
\ell & \ell' & L \\
m & m' & -M
\end{pmatrix}
f_{\ell\ell'L}^{EB}  
\left( \frac{C_{\ell'}^{EE}}{C_{\ell'}^{EE}+N_{\ell'}^{EE}} \hat{a}_{\ell'm'}^E \right)
\left( \frac{C_L^{\phi\phi}}{C_L^{\phi\phi}+N_L^{\phi\phi}} \hat{\phi}^{XY}_{LM} \right)=
\\ \equiv &
\left[ \hat{a}_{\ell'm'}^E \circledast \hat{\phi}^{XY}_{LM} \right]_{\ell m}.
\end{align}
This can then be used to subtract the lensing contribution from the total, measured B-mode field
\begin{equation}
    \hat{a}^{B\textrm{\ del.}}_{\ell m} = \hat{a}^B_{\ell m} -  \hat{a}^{B~\textrm{temp.}}_{\ell m},
\end{equation}
leaving a residual in the B-mode power spectrum given by \citep{Smith2012}
\begin{equation}
    C_\ell^{\textrm{\ res.}} = \frac{1}{2\ell+1} \sum_{\ell' L} \left| f^{EB}_{\ell\ell'L}\right|^2 \left[ C_{\ell'}^{EE} C_L^{\phi\phi} - \frac{\left(C_{\ell'}^{EE}\right)^2}{C_{\ell'}^{EE}+N_{\ell'}^{EE}} \frac{\left(C_L^{\phi\phi}\right)^2}{C_L^{\phi\phi}+N_L^{\phi\phi}}  \right]. \notag
\end{equation}
This leaves out additional terms when $\phi$ is a quadratic estimate of CMB fields. Further biases arise in the internally delensed B-mode power spectrum due to disconnected, Gaussian correlations in higher order n-point functions, as described for example in Refs.~\citep{Carron2017a,Namikawa2017a}. We account for this bias, the so-called delensing bias, $N_\ell^\textrm{del.}$, with Monte Carlo simulations including our purely Gaussian foreground model. This bias has two contributions, one from a four-point function and one from a six-point function of CMB fields
\begin{align}
    N_\ell^\textrm{del.} \equiv& \left\langle \left| \hat{a}^B -\left[ \hat{a}^E \circledast \left[ \hat{a}^X \ast \hat{a}^Y \right] \right] \right|^2 \right\rangle -C_\ell^\textrm{res.} \\=&
    - 2 \left\langle \hat{a}^{B\ \dagger} \left[\hat{a}^E \circledast \left[ \hat{a}^X \ast \hat{a}^Y \right]\right] \right\rangle+ \left\langle
    \left[ \hat{a}^E \circledast \left[ \hat{a}^X \ast \hat{a}^Y \right] \right]^\dagger \left[ \hat{a}^E \circledast \left[ \hat{a}^X \ast \hat{a}^Y \right] \right] \right\rangle -C_\ell^\textrm{res.},
\end{align}
where we consider the lensing potential to be reconstructed from the combination of CMB fields $X$ and $Y$.

\subsection{CMB Power Spectrum Estimation}

We use the pure pseudo-cross-spectrum approach to compute polarized CMB (cross-)power spectra of cut-sky CMB maps \citep{Smith2006,Grain2009}, implemented in the {\sc X2Pure} code.\\

The final, total B-mode power spectrum, $\hat{C}_\ell^{BB}$, that can be used to estimate $r$, is then modeled with the following separate components
\begin{equation}
    \hat{C}_\ell=r C_\ell^\textrm{prim.}  +C_\ell^\textrm{res.} + N_\ell + F_\ell^\textrm{res.} + N_\ell^\textrm{del.} + F_\ell^\textrm{del.},
    \label{eq:BBdecomposition}
\end{equation}
where $F_\ell^\textrm{del.}$ arises only due to non-Gaussian statistics of the foregrounds after template delensing. In this work we neglect the dependence of the delensing bias, $N_\ell^\textrm{del.}$, on $r$.

\section{Galactic Foreground Biases in the Quadratic Estimator before component separation}\label{sec:isolatingfgbias}

In this section we investigate the response of the lensing potential quadratic estimator to various foreground models within different observation boundaries. At this stage we only apply an adapted filter in harmonic space prior to lensing reconstruction which downweights modes with high foreground amplitude (see Sec.~\ref{sec:gaufilter} for a more detailed description) and no full foreground cleaning algorithm yet. We make use of the simulation pipeline of Sec.~\ref{sec:galsim} and show the resulting foreground-bias depending on statistical parameters describing the GMF and dust grains. We vary the fiducial parameters of the foreground model (see Tab.~\ref{tab:vansyngelparams}) within the ranges of the statistical uncertainties of the fit obtained in Ref.~\citep{Vansyngel2016} and estimate corresponding bounds of lensing power spectrum biases. We make use of the $1\sigma$-errors of the foreground model's parameters \citep{Vansyngel2016}, which were obtained from their fit to current Planck data within the Planck HFI 24\% mask.\\

\begin{table}
    \centering
    \begin{tabular}{cccccc}\toprule
        $l_0$ & $b_0$ & $p_0$ & $\alpha_M$ & $f_M$ & $N$ \\\midrule
        $70^\circ$ & $24^\circ$ & $0.25$ & $-2.5$  & $0.9$ & $4$\\\bottomrule
    \end{tabular}
    \caption{The fiducial parameters of the polarized dust emission simulations obtained by fitting simulations to Planck data in \protect\citep{Vansyngel2016}.}
    \label{tab:vansyngelparams}
\end{table}

\begin{figure}
\centering
\includegraphics[width=.66\linewidth]{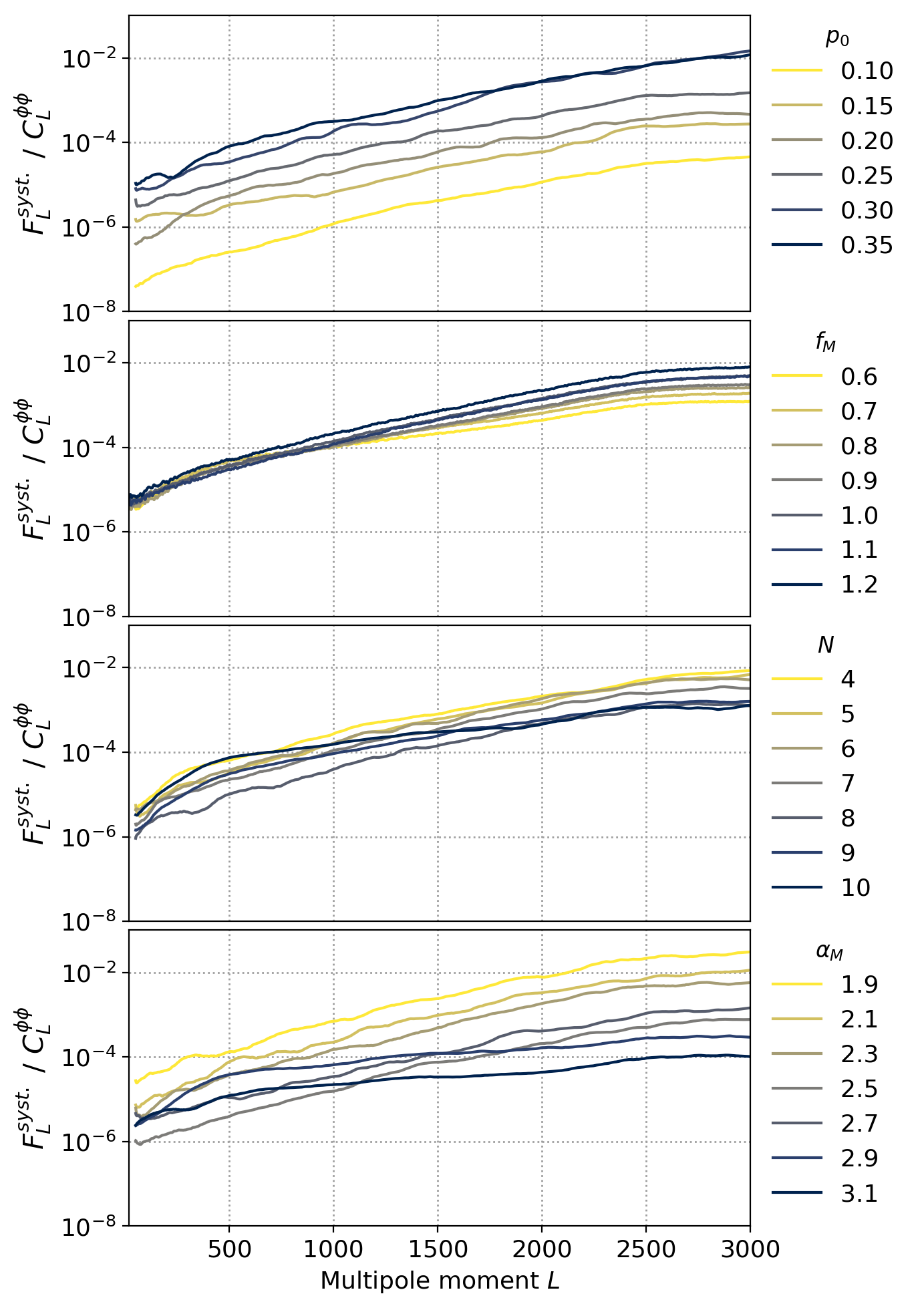}
\caption{The systematic bias, relative to the signal $C_L^{\phi\phi}$, in the lensing potential power spectrum estimate from polarization CMB data ($EBEB$) due to changing parameters of the galactic foreground model. We vary four parameters of the foreground model described in Sec.~\ref{sec:galsim}, $p_0$, $f_M$, $N$ and $\alpha_M$ independently, fixing the other parameters to the best-fit values of the fit of the model to the Planck data \citep{Vansyngel2016}. The shown parameter ranges stem from the corresponding $1\sigma$-uncertainties.}
\label{fig:biasp0}
\end{figure}

In Fig.~\ref{fig:biasp0} we show the behaviour of the resulting biases for the varying parameters $p_0$, $f_M$, $N$ and $\alpha_M$. As expected, the amplitude of the bias increases with $p_0$ and increasing complexity of the foreground model, either due to additional layers, $N$, or increasing the relative power of the turbulent galactic component, $f_M$. Further, increased small-scale correlations in the GMF, i.e. a lower spectral index, $\alpha_M$, leads to an increased amplitude of the bias. \\

\begin{figure}
\centering
\includegraphics[width=.8\linewidth]{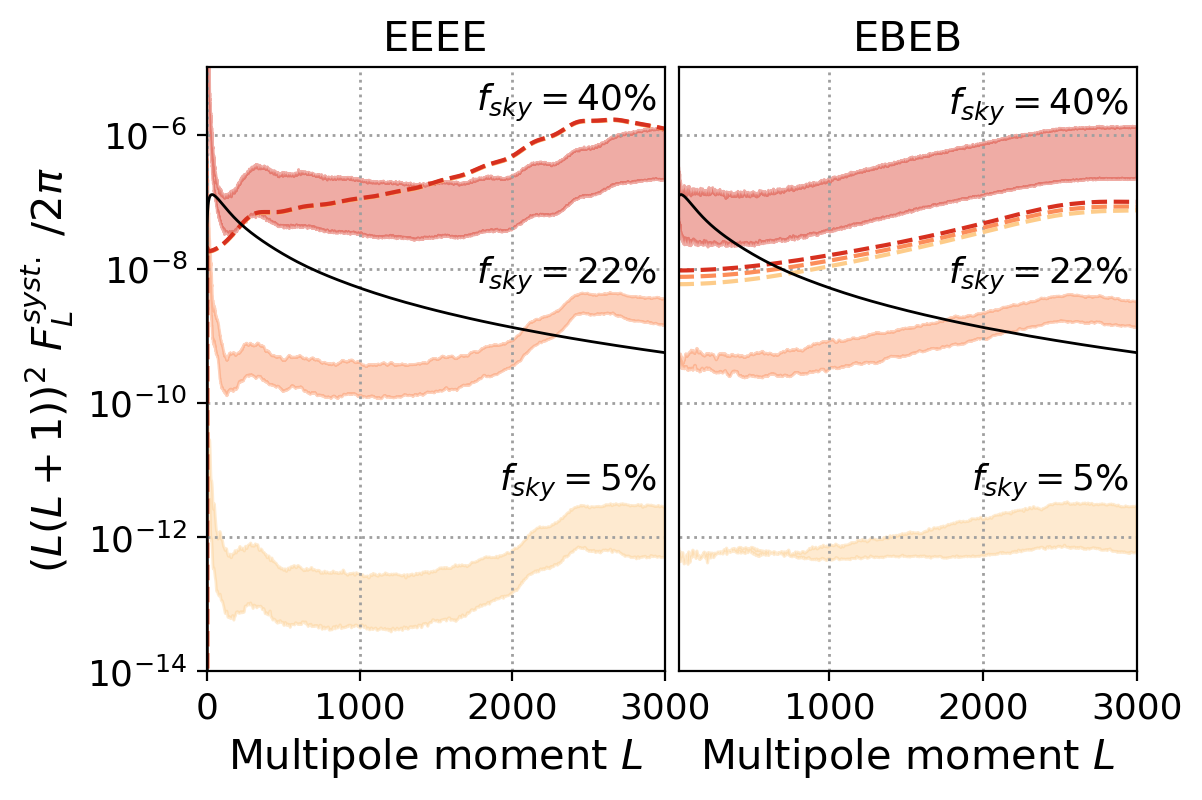}
\caption{The residual biases, $F_L^\textrm{syst.}$ of Eq.~\ref{eq:fgtrispectrum}, for a CMB-S4-like configuration on 5\%, 22\% or 40\% of the sky, measured in the $EEEE$ (left) and $EBEB$ (right) lensing potential power spectrum estimators. We produce foreground simulations for varying relative strength of the GMF turbulent component between $f_M=0.6$ and $f_M=1.2$, which lie within the $1\sigma$ uncertainty of the fit of \citep{Vansyngel2016}. The dashed lines show the respective $N^{(0)}$-bias for the respective sky fraction. The dashed curves on the r.h.s. overlap.}
\label{fig:biassnr}
\end{figure}

By considering the three survey strategies introduced in Fig.~\ref{fig:scansurveys}, we show in Fig.~\ref{fig:biassnr} the bias depending on the level of foreground contamination. The indicated bands correspond to varying levels of the turbulent component of the GMF, $f_M=0.9\pm0.3$. \\ 

\begin{figure}
\centering
\includegraphics[width=\linewidth]{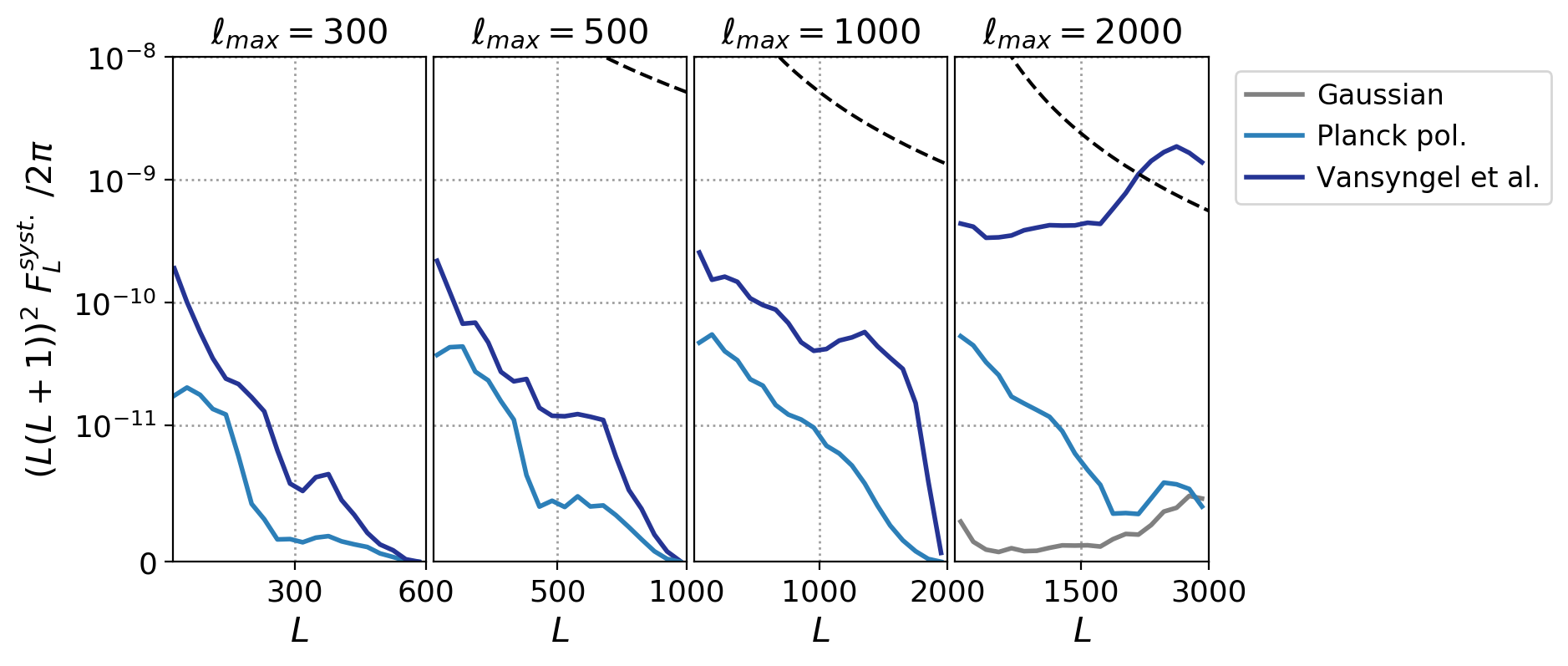}
\caption{Comparison of the biases in the $(EBEB)$ lensing potential power spectrum estimation from the \textsc{PySM} (Planck pol.) and Vansyngel et al. (Ref.~\citep{Vansyngel2016}) foreground models, respectively, as well as their dependence on the maximal multipole, $\ell_\textrm{max}$, considered in the reconstruction. The bias of the purely Gaussian foreground model obtained as a Gaussian realization of the fitted power spectra is denoted as (G) in the rightmost panel. The fiducial signal power spectrum, $C_L^{\phi\phi}$ is shown in the black, dashed line.}
\label{fig:biasmodels}
\end{figure}

Further we compare the resulting bias in the lensing power spectrum from different foreground models. In particular, we compare in Fig.~\ref{fig:biasmodels} the two dust models mentioned in Sec.~\ref{sec:galmodels}. On one hand the model of \textsc{PySM} relying on a dust and synchrotron emission template from Planck on signal-dominated scales until $\ell \approx 300$ and a power-law extrapolation and Gaussian realization at smaller scales. On the other hand the model of Ref.~\citep{Vansyngel2016} relying on GMF simulations. Fig.~\ref{fig:biasmodels} shows the evolution of the bias with varying scales $\ell_\textrm{max}$, above which the modes of the foreground simulations are set to zero. We observe that the level of the biases are similar at scales $\ell \approx 500$, while the majority of the power comes indeed from small scales which are supposedly modelled better in the GMF model. Note that the \textsc{PySM} model uses information from the measured polarization data of Planck, while the GMF model only uses intensity information. As a validation of our analysis, the same figure shows the level of the bias assuming a purely Gaussian foreground emission. Its level is negligible.

\section{Foreground Bias Mitigation}

\subsection{Gaussian and Isotropic Foregrounds}\label{sec:gaufilter}
\begin{figure*}
\centering
\includegraphics[width=\linewidth]{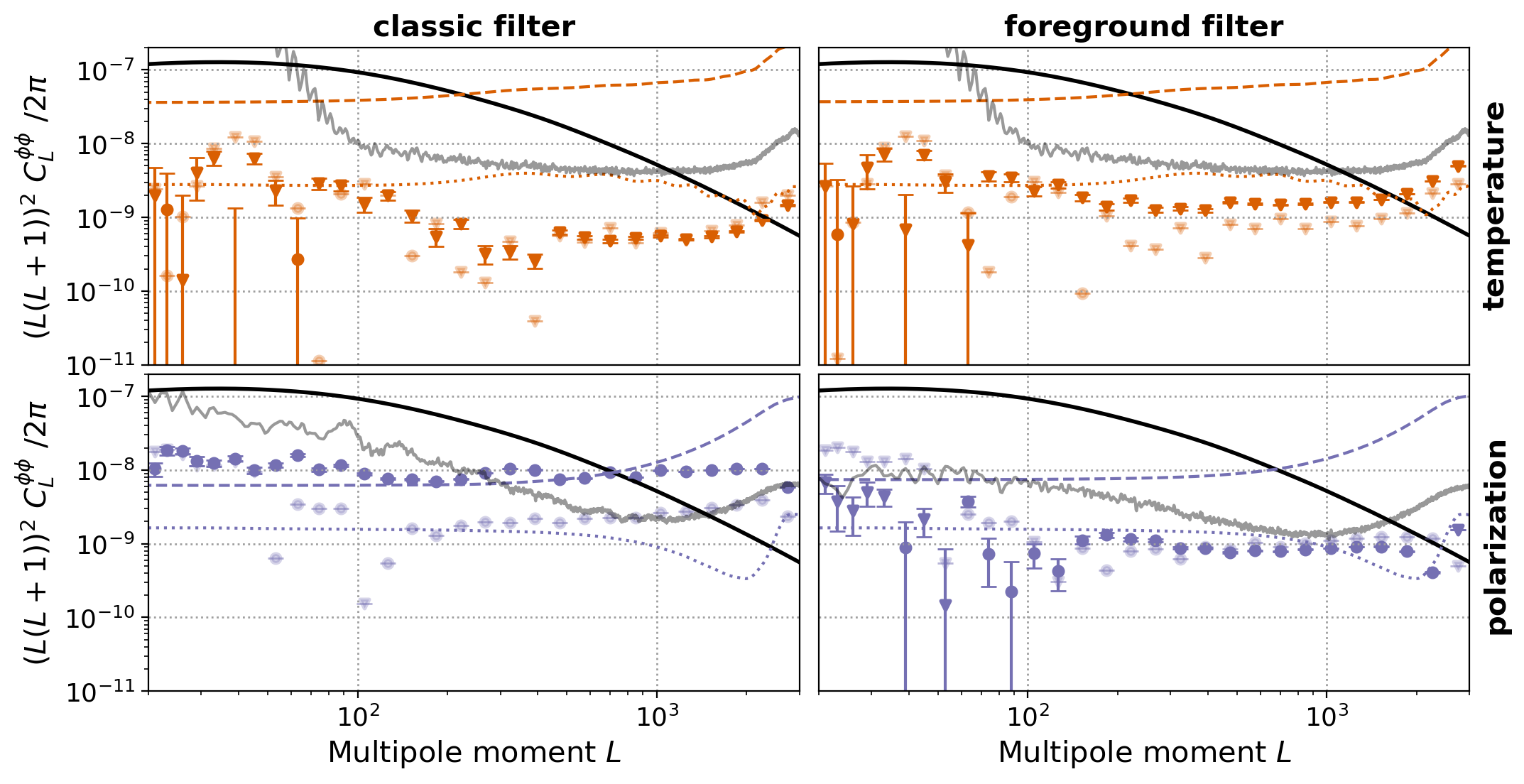}
\caption{The symbols depict are the biases, $F_L^\textrm{syst.}$ of Eq.~\ref{eq:fgtrispectrum}, of the lensing power spectrum after subtracting analytic $N^{(0)}$ and $N^{(1)}$ biases. We compare different weighting schemes of the quadratic estimator auto-power spectra of the temperature (top row) and polarization (bottom row) lensing estimations on the foreground simulations of \protect\citep{Vansyngel2016}, assuming a combination of the $95$ and $145\ \textrm{GHz}$ channel in our CMB-S4-like configurations for $5$\% (light colors) and $22$\% (bright colors) of the sky. The dashed and dotted lines show the analytic $N^{(0)}$ and $N^{(1)}$ biases, respectively, which are subtracted from the estimated spectra. On the left hand side the two-point contamination from foregrounds is not taken in account in the estimator's weights, on the right hand side we apply the substitution in Eq.~\ref{eq:replacecls}. In grey, we also show the mean-field power spectrum of the respective estimator on the bigger patch.}
\label{fig:filterbias}
\end{figure*}

To first order approximation, the galactic foreground emission can be described solely by its second moment, such that it is natural to absorb it within the weights and analytic normalization calculation of the quadratic estimator \citep{Challinor2018}
\begin{equation}
\hat{C}_{\ell}=\tilde{C}_{\ell}+N_{\ell}+F_{\ell}. \label{eq:replacecls}
\end{equation}
This maintains optimality of the quadratic estimator in the sense that it minimizes the resulting variance following the derivation in Ref.~\citep{Okamoto2003}. It can be seen as down-weighting the modes with respect to their foreground power and is in that sense less aggressive than the strategy proposed in Ref.~\citep{Fantaye2012} of introducing a high-pass filter to mitigate galactic foreground biases. In Fig.~\ref{fig:filterbias} we observe no bias-reducing effect for the temperature reconstruction, in which case the two-point spectrum is CMB dominated. For the polarization (EBEB) reconstruction, this filtering, effectively applying a low-pass filter on the polarization signal, is reducing the bias on scales up to $L=1000$. This filtering comes, however, with the caveat of increased noise in the final lensing power spectrum. Furthermore, the mean field for the $EB$ estimator gets reduced at the largest scales, reducing one simulation-dependent factor in the analysis well below the signal. 

\subsection{Component Separation}\label{sec:fgxlenscompsep}

In the context of component separation a second layer of complexity, the frequency scaling of the foregrounds, is added. The procedure is described in Sec.~\ref{sec:freqscalingsims}. We produce sets of multi-frequency maps from two foreground templates
\begin{itemize}
\item Gaussian foreground templates produced as a Gaussian realization of the power-law fit in Eq.~\ref{eq:fgpowerlaw}, scaled with a constant spectral index, given as the averaged spectral index over the considered sky region in the {\sc PySM} spectral index maps,
\item the simulations obtained by the method of Sec.~\ref{sec:galsim}, scaled with a line-of-sight varying spectral index and in the following denoted as {\it Vansyngel et al.}.
\end{itemize}

\begin{figure}
\centering
\includegraphics[width=.66\linewidth]{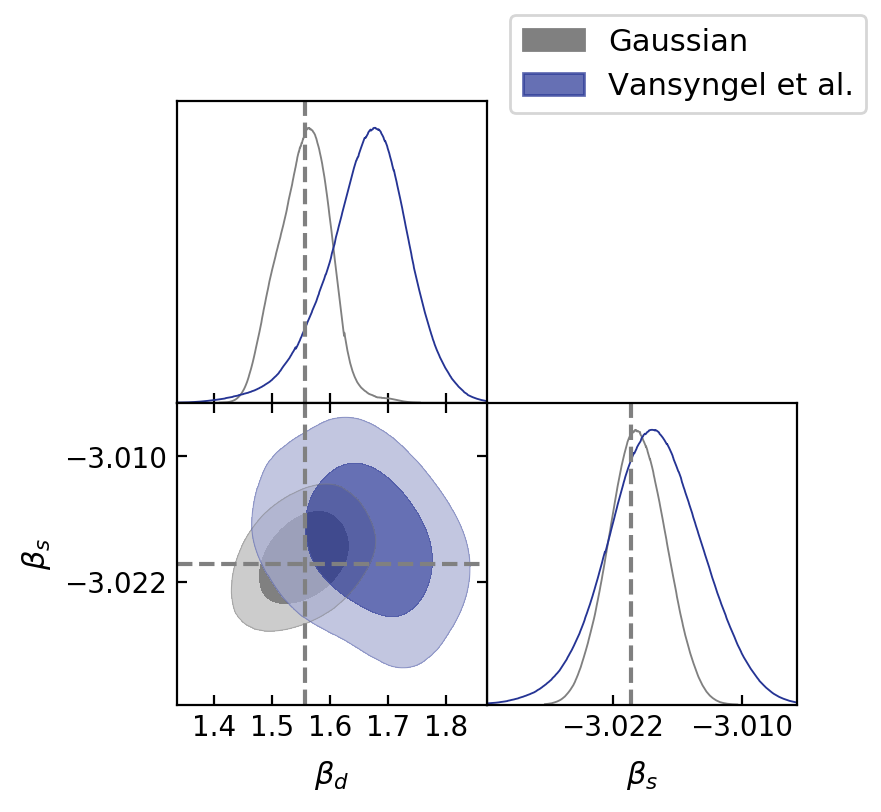}
\caption{Distributions of the best-fit foreground spectral parameters obtained by optimizing the spectral likelihood for $100$ CMB and noise simulations and two different foreground models. We compare the spectral parameters, $\beta_\textrm{d}$ and $\beta_\textrm{s}$, estimated on 5\% of the sky using the foreground models, purely Gaussian and Vansyngel at al. simulations, as described in Sec.~\ref{sec:fgxlenscompsep}. The dashed lines show the mean values, $\bar{\beta}_\textrm{d}$ and $\bar{\beta}_\textrm{s}$, over the patch.}
\label{fig:specparas_P}
\end{figure}

In Fig.\ \ref{fig:specparas_P} we show the result of the maximum-likelihood fit of the two cases of galactic foreground modeling on 100 simulations each, using routines of the \textsc{fgbuster} software package. In the fit we assume two foreground components, a modified black-body for dust and a power-law SED for synchrotron, with a single, spatially constant spectral index for both, $\beta_{\rm d}$ and $\beta_{\rm s}$, and a spatially constant dust temperature, $T_{\rm d}$. The grey, dashed horizontal lines mark the input value for the simulations with constant spectral indices. We recover the input values if the input and the assumed SED model coincide (in this case for the Gaussian foreground simulations), with slight systematic biases if that is not the case (e.g. the variation of the spectral indices in the Vansyngel et al.-simulations) for temperature are for illustration only.\\

\begin{figure}
\centering
\includegraphics[width=\linewidth]{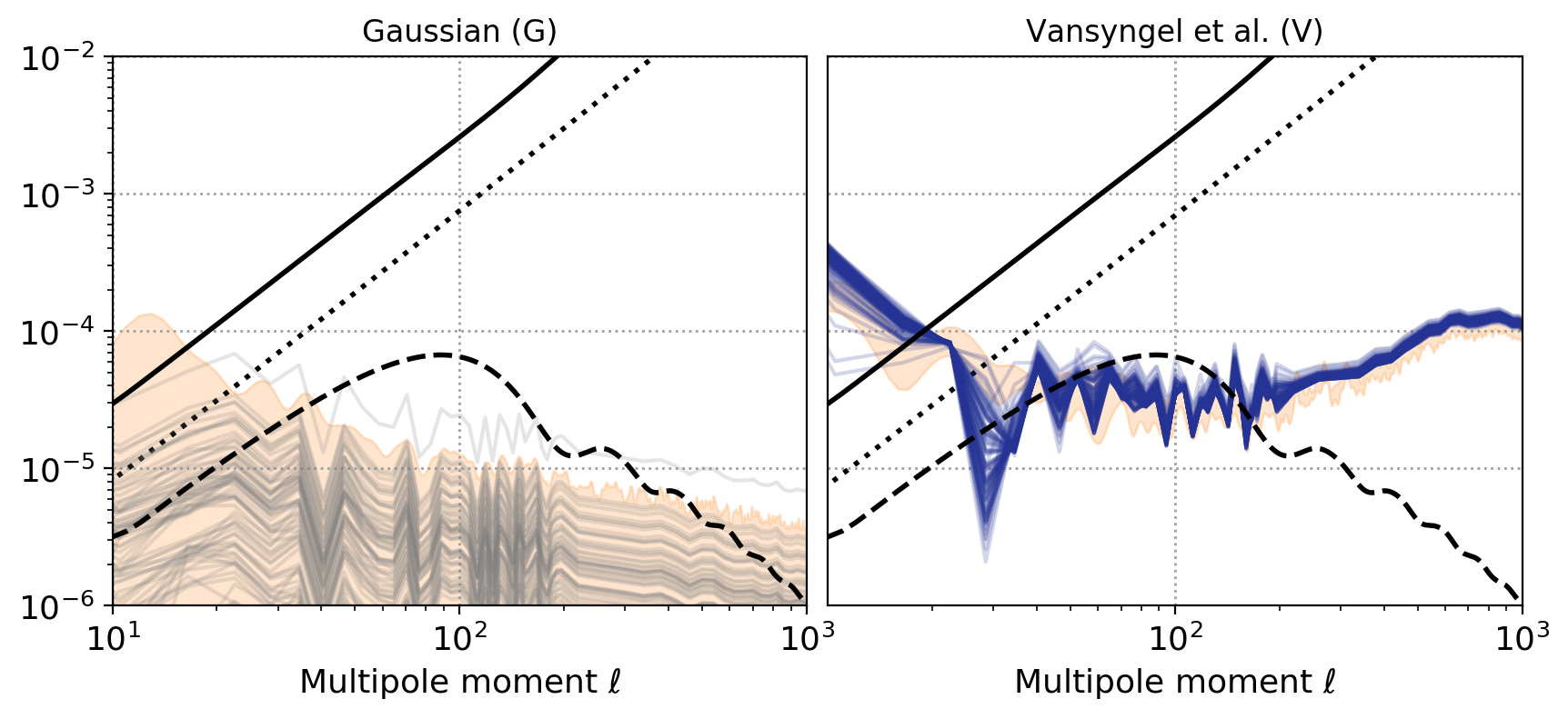}
\caption{The different components of the B-mode power spectrum after foreground cleaning for the two foreground models: the B-mode power spectrum due to lensing in the solid black line, the noise power after component separation in the dotted black line, the primordial B-mode power spectrum corresponding to our fiducial value of the tensor-to-scalar ratio, $r=10^{-3}$, in the dashed black line and the foreground residuals from our 100 simulations in the shaded grey or blue lines, respectively. The semi-analytically calculated $2\sigma$-scatter of the foreground residuals are shown in orange shade, cf. Eq. \ref{eq:2ptfgres}.}
\label{fig:fgresiduals}
\end{figure}

In Fig.\ \ref{fig:fgresiduals} we illustrate the foreground residuals as calculated for the B-mode auto-power spectrum obtained as described in Sec. \ref{sec:fgresidualcompuatation} using the best-fit values derived as described above. The scatter of the spectra in our 100 simulations illustrates the additional uncertainty, so-called statistical residuals, introduced in the B-mode power spectrum due to the uncertainty in the foreground SED estimations \citep{Stompor2016,Errard2018}. The simple propagation of the degraded noise properties after component separation to the Fisher forecast of the neutrino mass sensitivity from a CMB lensing potential measurement alone results in a $\approx 5$\% degradation for the temperature and a $\approx 25$\% degradation for the polarization estimator (see Tab.~\ref{tab:compseperror}), roughly the same for all considered sky patches.\\

\begin{figure*}[h]
\centering
\includegraphics[width=\linewidth]{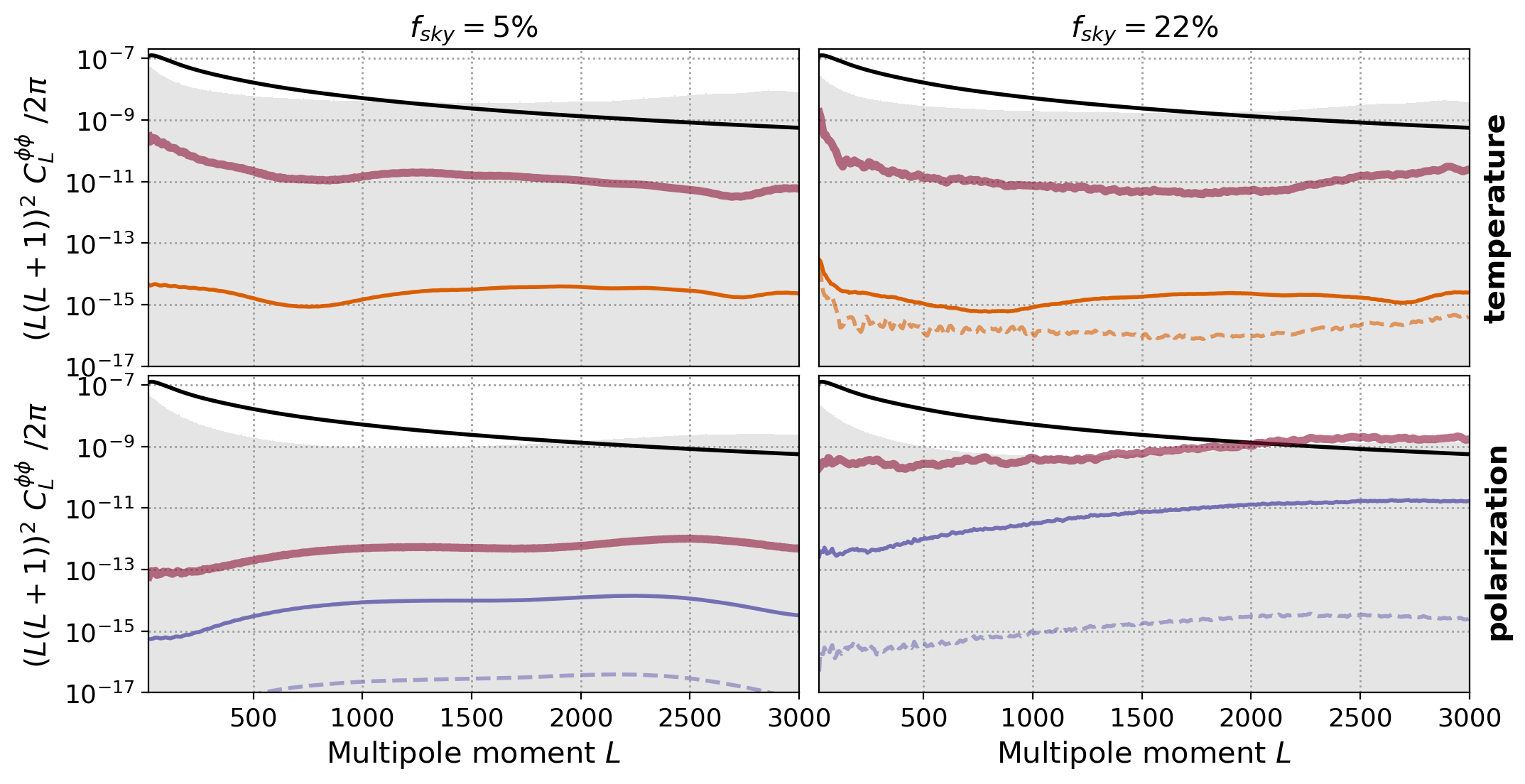}
\caption{The (absolute) bias in the CMB lensing power spectrum estimator before (red solid thick lines) and after (colored solid lines) component separation. The lensing potential estimation from temperature ($TTTT$) and polarization ($EBEB$) for the cases of observed sky fractions of $5$\% and $22$\% are shown. The fiducial lensing potential power spectrum is shown as a black solid line, the statistical uncertainty (after component separation) in each bandpower as grey bars. The difference in statistical uncertainty before and after component separation is negligible within the scaling of this plot. The corresponding power of the statistical foreground residuals, $F_L^\textrm{stat.}$, are shown as colored dashed lines in the respective panels.}
\label{fig:phiresbiascomp}
\end{figure*}

Analogous to Sec.~\ref{sec:gaufilter}, we include the power spectrum of the foreground residual in the CMB maps in the filtering of the quadratic estimator. We model the power spectra of systematic foreground residuals from single- and cross-frequency power spectra as 
\begin{equation}
\hat{C}_\ell^\textrm{res.}=\mathbf{W}_\ell^T \mathbf{F}_\ell \mathbf{W}_\ell,
\label{eq:residualpowermodel}
\end{equation}
where $\mathbf{F}_\ell$ is a $N_\textrm{freq.} \times  N_\textrm{freq.}$ matrix for each $\ell$, containing the multi-frequency foreground (cross)-spectra, obtained by fitting a power-law foreground spectrum to each (cross-)frequency power spectrum \citep{Stompor2016}. Fig.~\ref{fig:phiresbiascomp} shows the resulating biases due to diffuse foreground emission in the lensing potential power spectrum, $F_L^\textrm{syst.}$ and $F_L^\textrm{stat.}$, computed with the help of Eqs.~\ref{eq:fgtrispectrum} and \ref{eq:xfforlensing}, respectively. In the considered cases, small or large sky fractions and temperature or polarization lensing estimators, the systematic residuals dominate. The bias in both estimators is significantly reduced, by at least two orders of magnitude and are therefore well below the statistical uncertainty in our chosen bandpowers. In Tab.~\ref{tab:compseperror} we show subsequent forecasts of the total mass of neutrinos, $M_\nu$, performed by a simple, single-parameter Fisher-matrix formalism (e.g. as in \citep{Huterer2005}) from the CMB lensing power spectrum alone. We observe a significant bias of the total neutrino mass in the case of no foreground cleaning of the large-patch survey. After applying the simple foreground cleaning algorithm here reduces this bias well below the envisioned sensitivity of CMB-S4 while still controlling systematic residuals on a negligible level. We emphasize that the results for temperature are for illustration only, given that the adopted sky model is greatly oversimplified.

\begin{table}
\centering\footnotesize
\begin{tabular}{ll|cc|cc}\toprule
$\bm M_\nu$ \textbf{estimates} & $\bm f_\textrm{sky}$ & \multicolumn{2}{c|}{\textbf{before cleaning}} & \multicolumn{2}{c}{\textbf{after cleaning}} \\
 in meV & & Bias  & $1\sigma$ error & Bias & $1\sigma$ error
    \\ 
\midrule
\textbf{temperature}  & $5$\%   & $-4.6$ & $66.5$ & $0.0$  & $71.1$\\
                      & $22$\%  & $-2.8$ & $31.8$ & $0.0$  & $33.4$ \\\midrule
\textbf{polarization} & $5$\%   & $-0.1$ & $22.7$ & $0.0$ & $29.9$ \\
                      & $22$\%  & $-126$ & $11.9$ & $-0.6$ & $14.4$ \\\bottomrule
\end{tabular}
\caption{Forecasts of the neutrino mass bias and sensitivity given the CMB-S4-like experimental configurations including scales $L=2-3000$, estimated with a Fisher matrix formalism, comparing the bias and sensitivity of the temperature- or polarization-only lensing potential estimators before and after foreground mitigation by multi-frequency component separation. The two cases of $f_\textrm{sky}=5\%$ and $f_\textrm{sky}=22\%$ are shown. A fiducial neutrino of $M_\nu=60\textrm{ meV}$ is assumed. }
\label{tab:compseperror}
\end{table}

\section{Delensing}

\begin{figure}
\centering
\includegraphics[width=.66\linewidth]{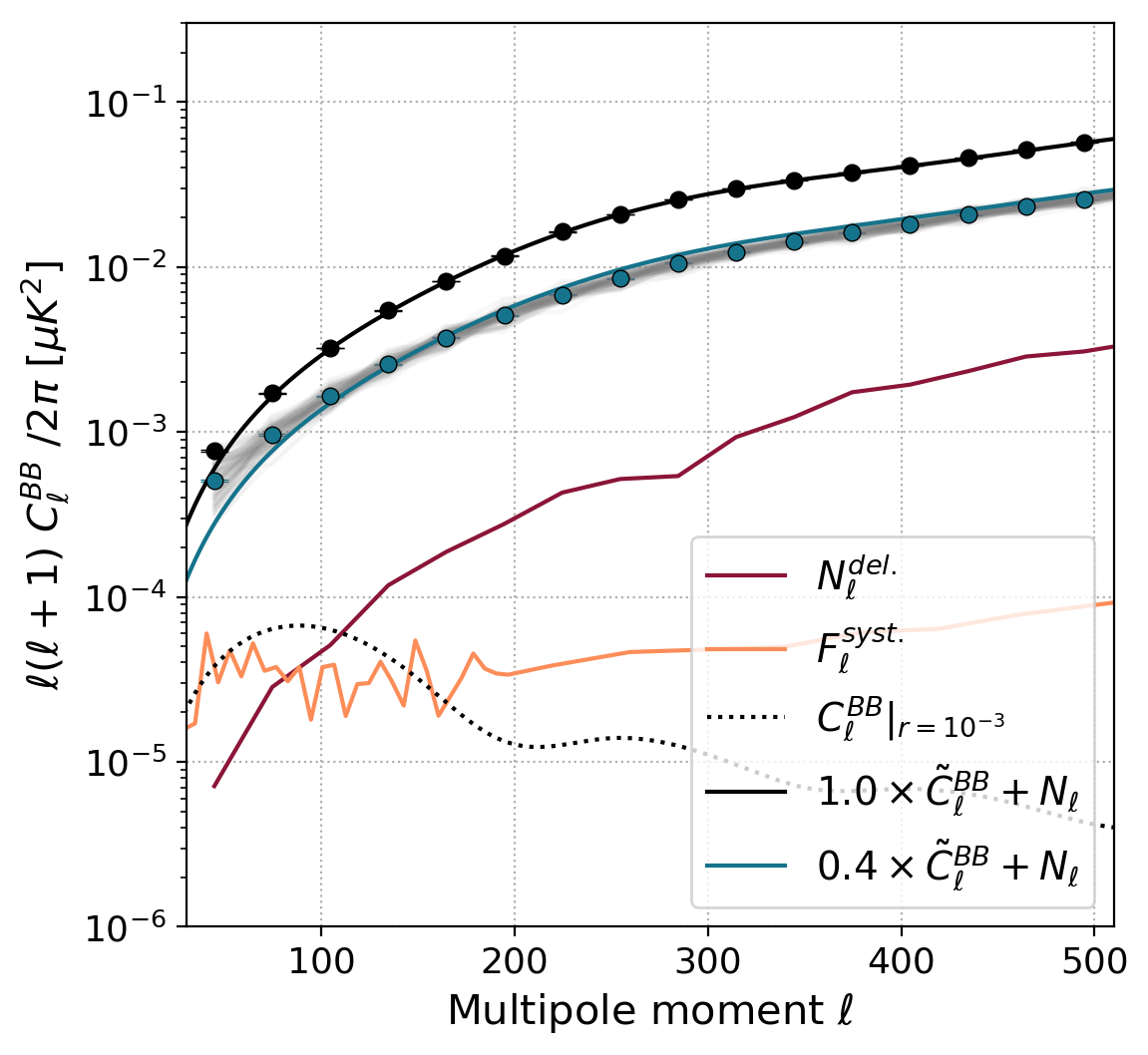}
\caption{An example of lensed ($\tilde{C}_\ell^{BB}$), delensed and primordial ($C_\ell^{BB}$) B-mode power spectra. The lines show the theory curves, where in solid black is the pure lensing power spectrum and in dotted black is the primordial B-mode power spectrum with $r=10^{-3}$. The blue line shows the lensed power spectrum, with a lensing amplitude of $A_\textrm{lens}=0.4$. The lensing power spectra include the instrumental noise bias. The red line shows the amplitude of the delensing bias, $-N_\ell^\textrm{del.}$. The filled circles show the mean of the $100$ simulations on the cleaned Vansyngel et al. simulation set (cf. Sec.~\ref{sec:galsim}), black for the lensed case, blue for the delensed case with the delensing and noise bias not removed. For the latter case the spectra from the $100$ realizations are shown in light grey. The foreground residuals, $F_\ell^\textrm{syst.}$, are shown in green.}
\label{fig:delexamp}
\end{figure}

\begin{figure}
\centering
\includegraphics[width=\linewidth]{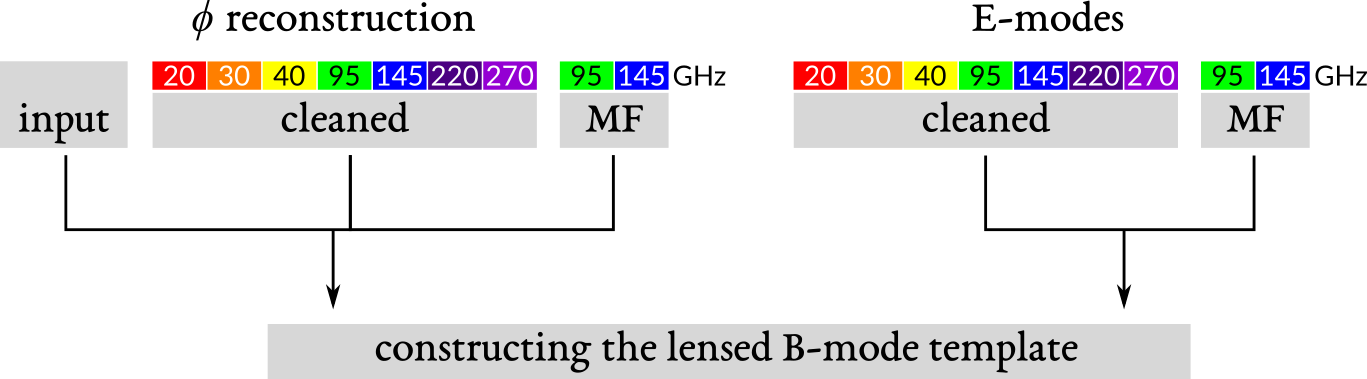}
\caption{Diagram summarizing the procedure to obtain delensed B-mode maps from various approaches of reconstructing the lensing potential, including the ideal case of the input $\phi$, combined with different E-mode estimators.}
\label{fig:deldiagram}
\end{figure}

To investigate foreground biases in the delensed B-mode power spectrum caused by Galactic foregrounds we produce two sets of simulations, one starting from Gaussian realizations of foreground power spectra and one starting from foreground templates with the method of Sec.~\ref{sec:galsim} (Vansyngel et al.). Both sets of multi-frequency simulations had been cleaned from foregrounds using the parametric method described earlier, leaving us with $100$ realizations of uncleaned and cleaned CMB $I$, $E$ and $B$ fields, as well as uncleaned and cleaned estimates of the lensing potential, $\phi$. These simulations can be used to estimate and isolate biases in the delensed B-mode power spectrum. As an example we show in Fig.~\ref{fig:delexamp} the various components of the B-mode power spectrum, comparing theory power spectra with the results from simulations. The simulated delensed spectra are consistent with a delensing amplitude of about $A_\textrm{lens}=0.4$, which is expected to be achievable for CMB-S4 with the quadratic estimator ($A_\textrm{lens}=0.1$ with an iterative delensing method \citep{Abazajian2016}). The delensing bias, $N_\ell^\textrm{del.}$, is computed using the simulations of purely Gaussian foregrounds, following Ref.~\citep{Namikawa2017a}. We identify the remaining bias after accounting for the foreground residuals (without any delensing), $F_\ell^\textrm{syst.}$, and the delensing bias (without any foreground cleaning) with the $F_\ell^\textrm{del.}$ of Eq.~\ref{eq:BBdecomposition}, which appears when treating foreground cleaning and delensing consistently.\\

\begin{figure}
\centering
\includegraphics[width=\linewidth]{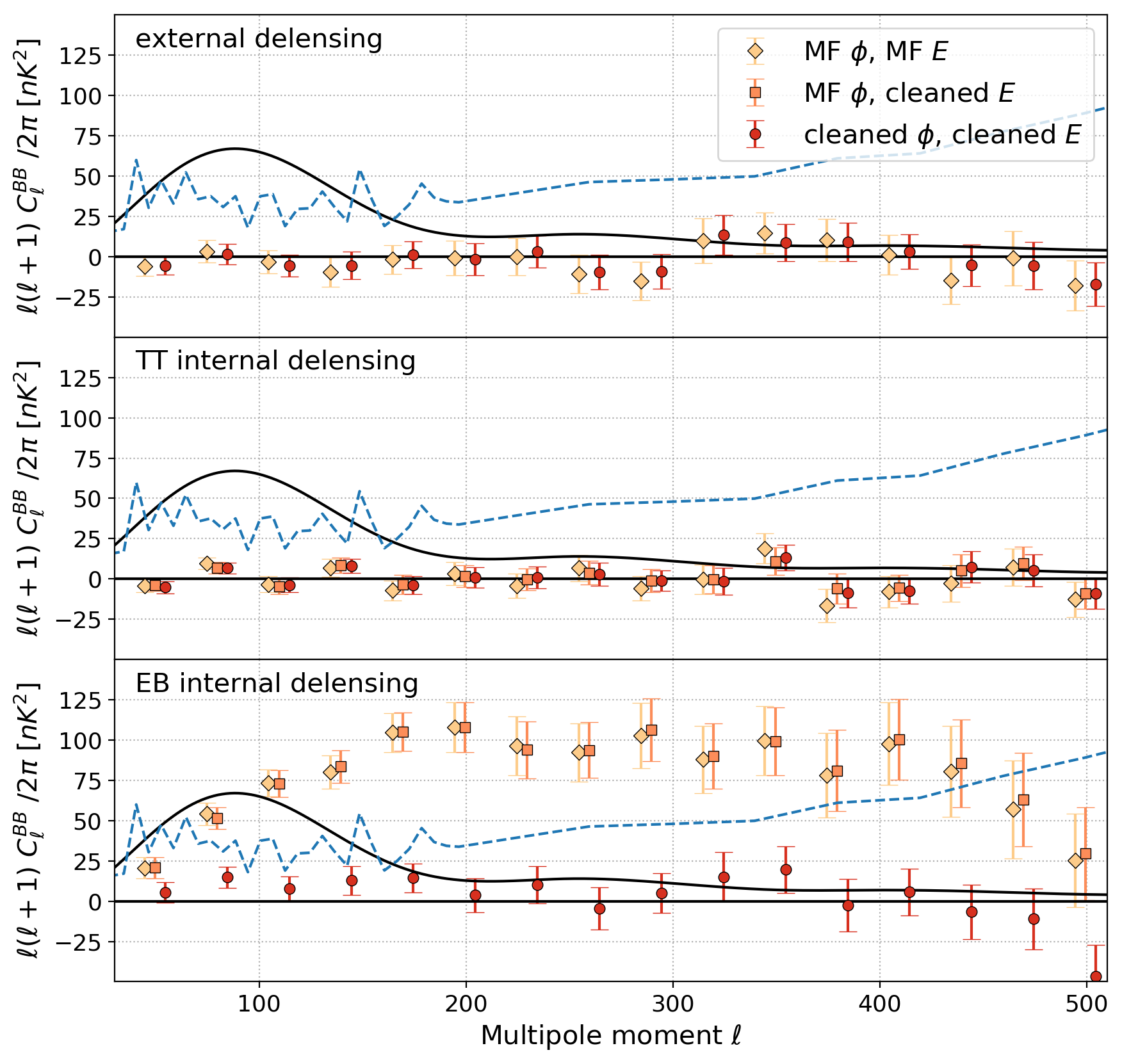}
\caption{Figure showing the $F_\ell^\textrm{del.}$-bias term of Eq.~\ref{eq:BBdecomposition}, originating from higher-order correlations of the diffuse foregrounds and computed from simulations, is shown as points. Contributions to the B-mode auto-power spectrum (in $nK^2$ units!) are shown. The dashed blue line is the average systematic foreground residual in the 5\% patch (same as the right hand figure of Fig.~\ref{fig:fgresiduals}). The solid black line is the primordial gravitational wave power spectrum, corresponding to $r=10^{-3}$. The two components of the B-mode template used for delensing, the $E$-mode field and the $\phi$ estimate can either be cleaned or not (called MF, medium frequency, i.e. $95$ and $145$ GHz only). }
\label{fig:bbias}
\end{figure}

Fig.~\ref{fig:deldiagram} depicts schematically the procedure to obtain lensed B-mode template maps. It also shows the various combinations of different experimental configurations we consider. We assume for both, lensing potential reconstruction and E-mode measurement, either a cleaned map or a simple coadded map from medium-frequency (MF) channels, a simple co-addition of the $95$ and $145$ GHz channels where the foreground contamination is not mitigated. We also compare this with the case where the lensing potential is not correlated with CMB fields by taking the input lensing potential. Fig.~\ref{fig:bbias} shows the resulting binned $F_\ell^\textrm{del.}$-bias, for three different cases of the B-mode template ingredients used for the delensing. We observe that external delensing, i.e. using the input $\phi$ potential as the delensing tracer, as well as $TT$ delensing, is free of this bias. When the lensing potential is estimated from $E$ and $B$ fields, higher-order correlations with the $E$ and $B$ fields that are used in the delensing procedure get introduced, causing a bias which is of the order of the primordial signal with $r=10^{-3}$ at its $\ell=80$-peak. It also effectively doubles the foreground residuals in the delensed map. However, this bias gets reduced if the LAT fields are cleaned from the foreground emission, i.e. both the $E$-mode field used for the B-mode template creation as well as the $E$ and $B$ fields used for the lensing tracer estimation.\\

We further propagate this bias to a likelihood on the tensor-to-scalar-ratio, $r$ \citep{Errard2016,Stompor2016,Carron2018}. Following Refs.~\citep{Tegmark1997,Hamimeche2008,Errard2018}, we employ a CMB-and-noise-averaged Gaussian likelihood on the CMB fields, which leads to
\begin{equation}
-2\log \mathcal{L} \left(r|\hat{C}_\ell^{BB}\right) = f_\textrm{sky} \left( \sum_\ell \frac{2\ell+1}{2} \frac{\hat{C}_\ell^{BB}}{C_\ell^{BB}} +\textrm{log} \left(\textrm{det}~ C_\ell^{BB}  \right) \right).
\end{equation}
The assumed covariance matrix is modeled including primordial and lensing contributions, as well as the noise and an estimate of the foreground residuals and Gaussian delensing biases (estimated from simulations including Gaussian foreground simulations)
\begin{equation}
C^{BB}_\ell(r)=r C_\ell^\textrm{prim.} +C_\ell^\textrm{res.} + N_\ell + F_\ell^\textrm{res.} + N_\ell^\textrm{del.},
\end{equation}
where off-diagonal contributions \citep{Schmittfull2013,Peloton2017} are neglected.\\

\begin{figure}
\centering
\includegraphics[width=.8\linewidth]{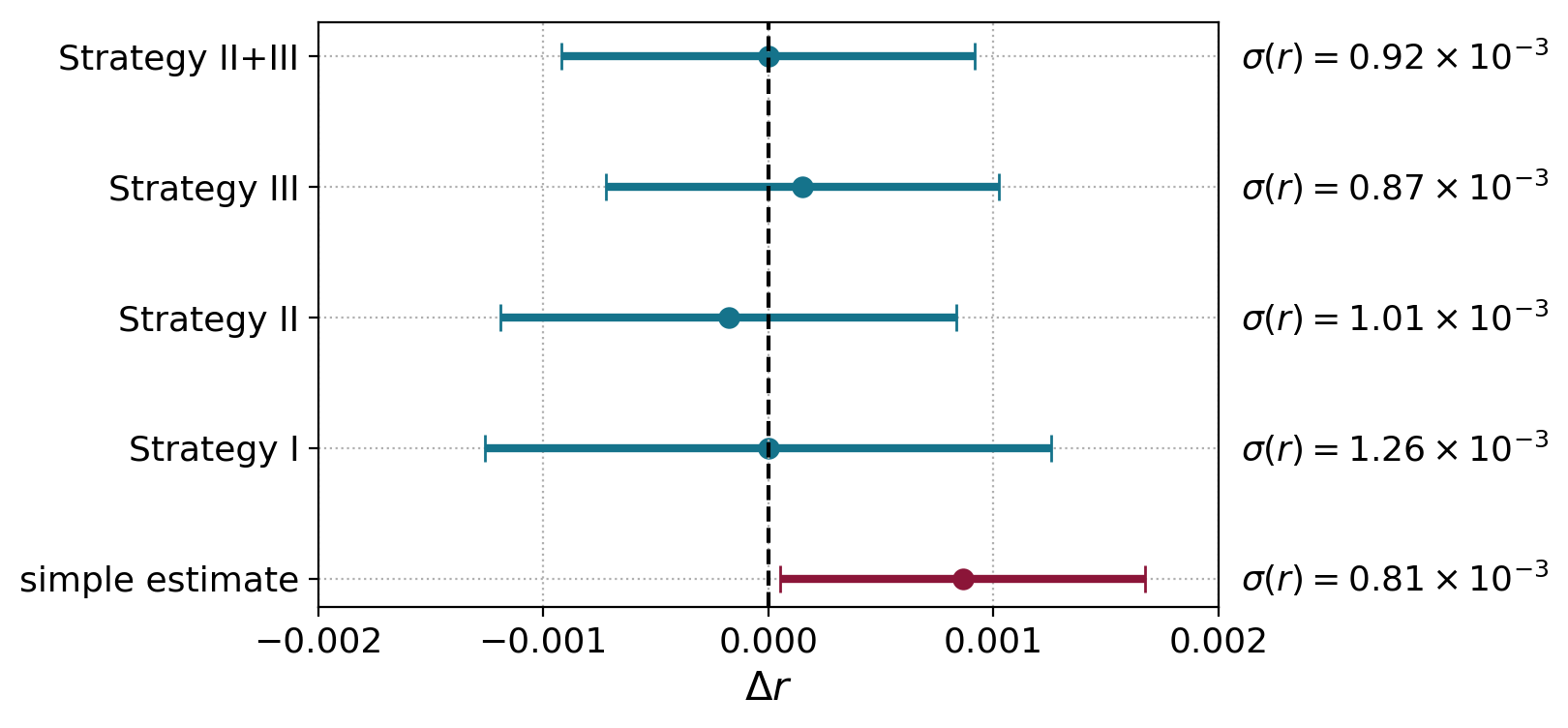}
\caption{From the maximum-likelihood fit on the B-mode power spectrum, we obtain a best-fit value and $1\sigma$ error estimate on $r$. We show a fiducial, simple case and three proposed bias mitigation strategies as proposed in the text.}
\label{fig:rbias}
\end{figure}

The biases and $1\sigma$-errors on $r$ for different configurations and reconstruction or delensing parameters are shown in Fig.~\ref{fig:rbias}. The fiducial value is $r=10^{-3}$. We include the B-mode auto-power spectrum between $\ell=30$ and $\ell=300$ in the $r$-likelihood. The measured B-mode map comes from foreground-cleaned SAT multi-frequency maps, with white noise specifications given in Tab.~\ref{tab:sensitivities} without additional atmospheric noise. We investigate three bias mitigation strategies:
\begin{itemize}
\item[I: ] Only using the internal lensing potential reconstruction from CMB temperature. We have seen that delensing with only the temperature quadratic estimator mitigates additional foreground biases, however, with the downside of having lower delensing efficiency.
\item[II: ] We remove multipoles of the CMB fields which are used for the lensing reconstructions below a certain $\ell_\textrm{min}$-value. This has the advantage that it reduces biases due to higher-order mode-mixing correlations \citep{Seljak2003,Teng2011,Sehgal2017}.
\item[III: ] We perform the galactic foreground cleaning technique introduced in the previous sections prior to lensing reconstruction and delensing.
\end{itemize}
Compared to the simple, fiducial case of performing no galactic foreground cleaning nor any other bias mitigation strategy prior to lensing reconstruction and delensing, which inhibits a $1\sigma$-bias to positive values, all mitigation strategies can remove the bias. The strategy which recovers the best sensitivity with a degradation by 7\% is a simple foreground cleaning technique. However, we show that even in the cases where this is not possible (i.e. there are no sufficient multi-frequency observations), the delensing estimator can otherwise be made more robust against biases from higher-order correlations of foregrounds, with a 25\% and 50\% degradation in the sensitivity of an $r$-measurement for Strategy I and II, respectively.

\section{Discussion}

We have investigated state-of-the-art small-scale diffuse foreground simulations, which can accurately reproduce one- and two-point statistics of dust and synchrotron emission data. We use those to investigate biases on the estimates of the CMB lensing potential arising from complex small-scale diffuse foregrounds. We find possible significant biases in large footprint surveys, as planned for Simons Observatory or CMB-S4. We explored a range of signal-to-noise levels of foreground biases, covering best- and worst-case scenarios in terms of the real realization of the magnetic field of our Galaxy, constrained by current Planck data. We claim that these biases can be mitigated by using multi-frequency observations and an application of standard foreground cleaning methods, given realistic galactic foreground simulations as described in Sec.~\ref{sec:galsim}. Furthermore, we tested for biases in the B-mode power spectrum, showing the necessity to clean the CMB of galactic foreground emission not only in dedicated large-scale CMB surveys (e.g. SATs), but also in small-scale measurements of the CMB (e.g. with LATs) to achieve the most sensitive and unbiased measurement of $r$. In the absence of any foreground cleaning significant biases in the $r$-estimation after internal delensing with the $EB$ quadratic estimator arise.\\

The GMF simulation method allows to produce extreme cases of statistical properties of the GMF by tuning the parameters of the simulation. Nonetheless, even in this case the turbulent component is simulated as a Gaussian random field, not supported by existing evidence provided by MHD simulations. This may indicate that the GMF simulations are still overly optimistic. We leave the inclusion of small-scale GMF simulations from MHD to future studies. Polarized small-scale galactic dust and synchrotron emission in the high signal-to-noise regime will be measured by next-generation CMB observatories such as Simons Observatory or CMB-S4. Also experiments like BLAST\footnote{\url{https://sites.northwestern.edu/blast/}}, S-PASS \citep{Carretti2019}, C-BASS\footnote{\url{https://cbass.web.ox.ac.uk}} and LiteBIRD \citep{matsumura14} will be critical in furthering our understanding of galactic foregrounds and characterizing possible biases in CMB lensing and delensing.

\acknowledgments

We thank Giulio Fabbian and Anthony Challinor for useful comments and discussions. We are grateful to Flavien Vansyngel for providing us galactic foreground simulations which were used in parts of this work. DB acknowledges support from Labex UnivEarthS. JE and RS acknowledge support of the French National Research Agency (Agence National de Recherche) grant, ANR BxB.

\bibliographystyle{JHEP}
\bibliography{refs.bib}

\end{document}